\documentclass[aps,pra,showpacs,superscriptaddress,tightenlines,twocolumn,lengthcheck]{revtex4-1}
\usepackage{diagbox}
\usepackage{multirow}
\usepackage{amsmath}
\usepackage{amsfonts}
\usepackage{graphicx}
\usepackage{epsfig}
\usepackage{color}
\usepackage{txfonts}
\usepackage[colorlinks,citecolor=blue]{hyperref}
\usepackage{diagbox}
\usepackage{mathrsfs}
\usepackage{xcolor}
\usepackage{amsfonts}
\usepackage{bbm}
\usepackage{dsfont}
\usepackage[T1]{fontenc}

\begin{document}
	
		\title{Dark-state engineering in Fock-state lattices}
		\author{Xuan Zhao}
		\affiliation{Key Laboratory of Low-Dimensional Quantum Structures and Quantum Control of Ministry of Education, Key Laboratory for Matter Microstructure and Function of Hunan Province, Department of Physics and Synergetic Innovation Center for Quantum Effects and Applications, Hunan Normal University, Changsha 410081, China}
		\affiliation{Hunan Research Center of the Basic Discipline for Quantum Effects and Quantum Technologies, Hunan Normal University, Changsha 410081, China}
		\author{Yi Xu}
		\affiliation{Key Laboratory of Low-Dimensional Quantum Structures and Quantum Control of Ministry of Education, Key Laboratory for Matter Microstructure and Function of Hunan Province, Department of Physics and Synergetic Innovation Center for Quantum Effects and Applications, Hunan Normal University, Changsha 410081, China}
		\affiliation{Hunan Research Center of the Basic Discipline for Quantum Effects and Quantum Technologies, Hunan Normal University, Changsha 410081, China}
		\author{Le-Man Kuang}
		%\email{lmkuang@hunnu.edu.cn}
		\affiliation{Key Laboratory of Low-Dimensional Quantum Structures and Quantum Control of
			Ministry of Education, Key Laboratory for Matter Microstructure and Function of Hunan Province, Department of Physics and Synergetic Innovation Center for Quantum Effects and Applications, Hunan Normal University, Changsha 410081, China}
		\author{Jie-Qiao Liao}
		\email{Corresponding author: jqliao@hunnu.edu.cn}
		\affiliation{Key Laboratory of Low-Dimensional Quantum Structures and Quantum Control of Ministry of Education, Key Laboratory for Matter Microstructure and Function of Hunan Province, Department of Physics and Synergetic Innovation Center for Quantum Effects and Applications, Hunan Normal University, Changsha 410081, China}
			\affiliation{Hunan Research Center of the Basic Discipline for Quantum Effects and Quantum Technologies, Hunan Normal University, Changsha 410081, China}
		\affiliation{Institute of Interdisciplinary Studies, Hunan Normal University, Changsha 410081, China}

	\begin{abstract}
			
		Fock-state lattices (FSLs) are becoming an emerging research hotspot in quantum physics, not only because the FSLs provide a new perspective for studying atom-field interactions, but also because they build a connection between quantum optics and condensed matter physics. Because of the multiple transition paths in the lattices, an inherent quantum interference effect exists in these systems, and hence finding new quantum-coherent phenomena and exploiting their applications become a significant and desired task in this field. 
		In this work, we study the dark-state effect in the FSLs generated by the multimode Jaynes-Cummings (JC) models. By considering the FSLs in certain excitation-number subspaces, we study the dark states with respect to the states associated with the atomic excited state using the arrowhead-matrix method. We find that dark-state subspaces exist with dimensions determined by the number of orthogonal dark states. When the dimension is greater than one, the forms of these dark-state bases are not unique. Further, we obtain the number and form of the orthogonal dark states in the two-, three-, and four-mode JC models. In addition, we find that for a general $N$-mode JC model, there are $C_{N+n-2}^{N-2}$ orthogonal dark states in the $n$-excitation subspace. We also build the relationship between the dark states and dark modes. Our work paves the way for exploring quantum optical effects and quantum information processing based on the FSLs.

	\end{abstract}
	
	\date{\today}
	\maketitle
	
	\section{Introduction}
	
	The lattice is an important concept introduced to understand the structures and properties of crystals, and it plays a crucial role in the development of solid-state physics and condensed matter physics~\cite{SSP:M1976}. While traditional lattices have achieved great success in describing various novel physical phenomena~\cite{PToLD:B1983}, they are not flexible or controllable enough because of the inherent structure of materials, which leads to some limitations in quantum manipulations of matter.
	Fock-state lattices (FSLs)~\cite{FSL:PRL2016,TP:NC2021,FSL:SL2021,topology:science2022,FSLtQO2023,QSiFSL2024}, as a new type of quantum structure, have attracted much attention from the peers of quantum physics and condensed matter physics. The FSLs offer flexibility in design and implementation because of the tunability of their size and geometry~\cite{FSL:PRL2016,FSL:SL2021}.
	Recently, the FSLs have been widely studied in coupled atom-field models. For example, researchers have studied the FSLs in multimode Jaynes-Cummings (JC) models~\cite{TP:NC2021,topology:science2022,QSiFSL2024}. It has been found that the one-, two-, and three-dimensional lattices can be, respectively, constructed in the two-, three-, and four-mode JC models, and that the topological physics can be studied in these FSLs. In addition, the FSLs have been studied in other physical models in quantum optics~\cite{FSLtQO2023}. 
	From the view-point of theoretical research, the physical properties of the FSLs are worthy of attention and deserve study.
	
	The dark-state effect~\cite{DS1976,ScullyQO1997}, as a coherent physical effect induced by quantum interference cancellation, has significance in both fundamental quantum physics and modern quantum optical technology~\cite{DSicqed2009,DSiJC2013,DS_DSp2014,DS:jpa2014,DS_DSp2016,DS:jpa2017,DS:prb2020,DS:jpl2021,DS:jpa2022,DS:CIarX,DS:prx2022}. 
	Many physical effects associated with the dark states, such as coherent population trapping~\cite{CPT1978,CPT1982,CPT1988,CPT1996,CPT1998} and electromagnetically induced transparency~\cite{EITPRL1991,Arimondo1996,EIT:H1997,DPiEIT2000,EIT2005}, have been extensively studied. 
	In particular, the dark states have been widely used in the implementation of stimulated Raman adiabatic passage (STIRAP)~\cite{STIRAP:A1989,STIRAP2015,STRIP2017}, which is a very useful mechanism for realizing quantum state transfer~\cite{QSTiQN1997,STIRAP2016,STRIP:sf2020} and quantum frequency conversion~\cite{FC2013,FC2014}. Recently, much attention has been paid to the study of dark states in coupled atom-field systems~\cite{DS_AF2020,DS_AF2021,DS_AF2022,DS_AF2023,DS_AF2024,CQED2024}. In addition, the concept of dark states is extended to the dark mode~\cite{DM2013}, and the dark-mode effect in networks has been studied~\cite{ODM2012,HFQST2012,ASC2012,huang2023dark}. In particular, a general method for determining the number and form of orthogonal dark modes in networks has been proposed~\cite{huang2023dark}. Since the FSLs can be understood as networks, in which the nodes and the links are, respectively, implemented by Fock states and transition channels, it is an interesting and important topic to study the dark-state effect in FSLs.
	
	In this work, we consider the multimode JC models~\cite{JC1963,JCmodel1993,JCandD2022}, which describe the couplings between a two-level atom (TLA) and multimode bosonic fields. We create the FSLs of the system in the subspaces with a fixed number of excitations.
	Concretely, we analyze the dark states in the multimode JC models using the arrowhead-matrix method~\cite{huang2023dark}. By defining the upper states (the states associated with the atomic excited state) and the lower states (the states associated with the atomic ground state), the system basis states can be divided into two subcomponents. After defining the basis vectors for specific excitation-number subspaces, we can express the multimode JC Hamiltonian as a matrix $ \{ \{\mathbf{U},\mathbf{C}\}, \{ \mathbf{C}^{\dag} , \mathbf{L} \}\} $, where $\mathbf{U}$ and $\mathbf{L}$ are, respectively, the submatrices corresponding to the upper- and lower-state components, and $\mathbf{C}$ is the coupling matrix describing the couplings between the two subcomponents of states. When there exist degenerate lower states, the number and form of the dark states can be obtained by analyzing the coupling matrix $\mathbf{C}$ (the submatrix corresponding to the degenerate part). Using the matrix elementary transformation, we can transform a certain column into a zero vector, namely, the basis vector corresponding to this zero-vector column is decoupled from all these upper states. Therefore, the superposed lower states associated with the zero vector become the dark state. When there exist multiple degenerate dark states, these dark states form a dark-state subspace. All the states in the dark-state subspace are decoupled from all the upper states. The dimension of this dark-state subspace is determined by the number of orthogonal dark states, which can be used as the basis of the dark-state subspace. It should be noted that the forms of these orthogonal dark states are not unique. Using the above-mentioned method, we obtain the number and form of the orthogonal dark states for the two-, three-, and four-mode JC models in arbitrary-excitation subspace.
	For a general $N$-mode JC model, we present the number of orthogonal dark states in an arbitrary $n$-excitation subspace. 	In addition, we study the dark-mode effect in the multimode JC models. We also build the connection between the dark states and dark modes.
	
	The rest of this paper is organized as follows. In Sec.~\ref{sec2}, we introduce the $N$-mode JC model and present the general method for analyzing the dark-state effect in the FSLs. In Secs.~\ref{sec3}-~\ref{sec5}, we generate the FSLs in different excitation subspaces and analyze the dark states for two-, three-, and four-mode JC models. In particular, we obtain the general form for the orthogonal dark states. We also establish the connection between the dark states and the dark modes. In Sec.~\ref{sec6}, we obtain the number of dark states in the FSLs for a general $N$-mode JC model. Finally, we conclude this work in Sec.~\ref{conclusion}.
	 
	\section{The multimode JC models and the arrowhead-matrix method}\label{sec2}

\begin{figure}[t!]
		\centering\includegraphics[width=0.48\textwidth]{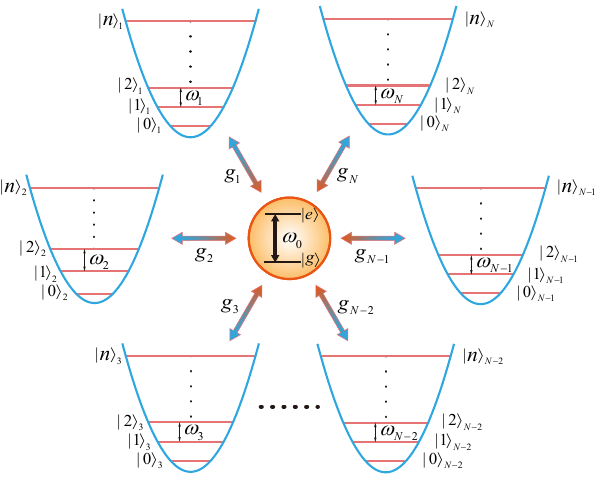}
		\caption{Schematic of the multimode JC model composed of a TLA with the energy
			separation $\protect\omega _{0} $ (between the excited state $\left\vert e\right\rangle$ and the ground state $\left\vert g\right\rangle$) and $N$ quantum field modes $a _{j=1-N}$ with the resonance frequencies $\omega _{j}$ and basis states $\left\vert n\right\rangle_{j}$. The variable $g_{j}$ denotes the coupling strength between the TLA and the $j$th field mode $a_{j}$.}
		\label{model}
	\end{figure}

	We consider an $N$-mode JC model, which describes a TLA coupled to $N$ quantum field modes (see Fig.~\ref{model}). The Hamiltonian of the $N$-mode JC model reads (with $\hbar =1$) 
	\begin{equation}
		H_{[N]}=\frac{\omega _{0}}{2}\sigma _{z}+\sum_{j=1}^{N}[\omega
		_{j}a_{j}^{\dag }a_{j}+g_{j}(\sigma _{-}a_{j}^{\dag }+a_{j}\sigma _{+})],
		\label{Hamiltonian}
	\end{equation}%
	where $\omega_{0}$ is the energy separation between the excited state $\left\vert e\right\rangle $ and ground state $\left\vert g\right\rangle $ of the TLA, which is described by the Pauli operators  $\sigma _{x}=\left\vert e\right\rangle
	\left\langle g\right\vert +\left\vert g\right\rangle \left\langle
	e\right\vert $, $\sigma _{y}=i(\left\vert g\right\rangle \left\langle
	e\right\vert -\left\vert e\right\rangle \left\langle g\right\vert )$, and $\sigma _{z}=\left\vert
	e\right\rangle \left\langle e\right\vert -\left\vert g\right\rangle
	\left\langle g\right\vert $. The raising and lowering
	operators are defined by $\sigma _{\pm }=(\sigma _{x}\pm i\sigma _{y})/2$.
	The $\omega_{j}$ is the resonance frequency of the $j$th field mode described by the annihilation (creation) operator $a_{j}$ $(a_{j}^{\dag })$. The parameter $g_{j}$ is the JC coupling strength between the TLA and the $j$th field mode $a_{j}$, and we assume real-valued $g_{j}$ throughout this work. 
	
    For better analyzing the dark-state effect, we work in the rotating frame defined by the unitary operator $U=\text{exp}[-\mathrm{i}\omega_{0}t(\sigma _{z}/{2}+\sum_{j=1}^{N}a_{j}^{\dag }a_{j})]$, then the Hamiltonian in Eq.~(\ref{Hamiltonian}) becomes%
	\begin{equation}
		\tilde{H}_{[N]}=\sum_{j=1}^{N}[-\Delta _{j}a_{j}^{\dag
		}a_{j}+g_{j}(\sigma _{-}a_{j}^{\dag }+a_{j}\sigma _{+})], \label{Hamiltonian2}
	\end{equation}%
	where we introduce the detuning $\Delta _{j}=\omega _{0}-\omega _{j}$ between the TLA and the $j$th field mode. In this model, the total excitation number operator, defined by $\mathrm{\hat{N}}_{[N]}=\left\vert e\right\rangle \left\langle e\right\vert+\sum_{j=1}^{N}a_{j}^{\dag
	}a_{j} $, is a conserved quantity because $[H_{[N]},\mathrm{\hat{N}}_{[N]}]=0$.
	As a result, below we discuss the dark states of the FSLs in different excitation-number subspaces.
	
	For the multimode JC model, we construct the FSLs by taking the Fock states as nodes and the transition channels between nodes as links. The lattices can be described by matrices in the state space, and the diagonal and off-diagonal matrix elements represent the on-site energies and transition amplitudes between different Fock states, respectively. 
	We can also understand the FSLs as networks. The connectivity of the networks is related to the transition chain, and these transitions are fundamentally determined by the symmetry of the system. Therefore, the construction of FSLs is necessary and it plays a substantive role in the dark-state effect. This is because the dark state is caused by quantum destructive interference between transition paths. 
	Moreover, the FSLs provide an intuitive and effective way to depict the transition structure among quantum states in the basis-state space, thereby offering a clear physical picture for analyzing the dark states. The method of visualizing coupling is also used in other physical scenarios~\cite{visual2009}. Furthermore, the FSLs naturally scale with the photon number, enabling high-dimensional expansion, and their geometry structure can be adjusted by designing the couplings and drivings.
	For a given FSL, its Hamiltonian can be expressed as a thick arrowhead matrix~\cite{huang2023dark}. 
	
    From the perspective of matrices, the dark-state problem can be reformulated as a null space problem of the coupling matrix, which simplifies calculations and improves efficiency.
	In this paper, we employ the arrowhead-matrix method to study the dark-state effect in the FSLs. 
	
	Depending on whether the atom is in the excited state or ground state, we can define the upper and lower states. Then the basis states of the system can be divided into two subcomponents.
	The upper states take the form $\left\vert e,n_{1}^{\prime},n_{2}^{\prime},...,n_{N}^{\prime} \right\rangle$ with $\sum_{j=1}^{N}n_{j}^{\prime}=n-1$ for natural numbers $n_{j}^{\prime}$, where the TLA is in the excited state and $n-1$ photons are placed in the $N$ field modes. The distribution of these $n-1$ photons in $N$ modes can be described by a classical permutation and combination problem, i.e., putting $n-1$ balls in $N$ boxes. Then there are $C_{N+n-2}^{N-1} $ possible arrangements for the corresponding photon number distributions.
	The lower states take the form $\left\vert g,n_{1},n_{2},...,n_{N}\right\rangle$ with $\sum_{j=1}^{N}n_{j}=n$ for natural numbers $n_{j}$. The distribution of the lower states is equivalent to placing $n$ balls in $N$ boxes, and there are $C_{N+n-1}^{N-1}$ possible arrangements. 
	We sort these states as {$\left\vert e,n-1,0,0,...,0\right\rangle$, $\left\vert e,n-2,1,0,...,0\right\rangle$, $\left\vert e,n-2,0,1,...,0\right\rangle$, ..., $\left\vert e,n-1-s_{2}^{\prime},s_{2}^{\prime}-s_{3}^{\prime},...,s_{n-1}^{\prime}-s_{n}^{\prime},s_{n}^{\prime}\right\rangle,$ ..., $\left\vert e,0,0,0,...,n-1\right\rangle$, $\left\vert g,n,0,0,...,0\right\rangle$, $\left\vert g,n-1,1,0,...,0\right\rangle$, $\left\vert g,n-1,0,1,...,0\right\rangle$, ..., $\left\vert g,n-s_{2},s_{2}-s_{3},...,s_{n-1}-s_{n},s_{n}\right\rangle$, ..., $\left\vert g,0,0,0,...,n\right\rangle$\} with $0\leq s^{\prime}_{n} \le s^{\prime}_{n-1} \leq ...\leq s^{\prime}_{3}\leq s^{\prime}_{2}\leq n-1$, and $0\leq s_{n} \le s_{n-1} \leq ...\leq s_{3}\leq s_{2}\leq n$. Then by defining the basis vectors corresponding to these basis states as $( 1,0,0,0,...0) ^{T}$, $(0,1,0,0,...0) ^{T}$, ..., and $( 0,0,0,...0,1) ^{T}$, the matrix corresponding to the Hamiltonian restricted in the $n$-excitation subspace can be written as
    \begin{equation}
    	\tilde{H}_{[N]}^{(n)}=\left( 
    	\begin{array}{c|c}
    		\mathbf{U}_{[N]}^{(n)} & \mathbf{C}_{[N]}^{(n)} \\ \hline
    		\left(\mathbf{C}_{[N]}^{ (n)}\right)^{\dag } & \mathbf{L}_{[N]}^{(n)}
    	\end{array}
    	\right) ,\label{HNn}
    \end{equation}%
    where $\mathbf{U}_{[N]}^{(n)}$, $\mathbf{L}_{[N]}^{(n)}$, and $\mathbf{C}_{[N]}^{(n)}$ are the submatrices related to the upper-state component, lower-state component, and the couplings between the two subcomponents of states, respectively. Note that the superscript \textquotedblleft$(n)$\textquotedblright\ and the subscript \textquotedblleft$[N]$\textquotedblright\ are used to denote the excitation number associated with the subspace and the mode number of the fields in the multimode JC model, respectively.
    
    All these matrix elements in Eq.~(\ref{HNn}) can be calculated based on the given basis vectors and the Hamiltonian. Based on the Hamiltonian in Eq.~(\ref{Hamiltonian2}), there exist couplings between the atom and field modes, while no interactions between these fields. Since there are no photon-hopping interactions between these field modes, the two submatrices $\mathbf{U}_{[N]}^{(n)}$ and $\mathbf{L}_{[N]}^{(n)}$ are diagonal. Moreover, based on the Hamiltonian in Eq.~(\ref{Hamiltonian2}), we can obtain the expression
    \begin{eqnarray}
    	&&\tilde{H}_{[N]}^{(n)}\left\vert e,n_{1}^{\prime },n_{2}^{\prime
    	},...,n_{N}^{\prime }\right\rangle \notag  \\
    	&=&\sum_{j=1}^{N}(-\Delta _{j}n_{j}^{\prime })\left\vert e,n_{1}^{\prime
    	},n_{2}^{\prime },...,n_{N}^{\prime }\right\rangle \notag \\
    	&&+\sum_{j=1}^{N}g_{j}\sqrt{n_{j}^{\prime }+1}\left\vert g,n_{1}^{\prime
    	},n_{2}^{\prime },...,n_{j}^{\prime }+1,...,n_{N}^{\prime }\right\rangle ,
    \end{eqnarray}%
     which indicates that each row of the matrix $\mathbf{C}_{[N]}^{(n)}$ contains $N$ nonzero matrix elements. When there exist degenerate lower states, the number and form of the dark states can be obtained by analyzing the coupling matrix $\mathbf{C}_{[N]}^{(n)}$~\cite{huang2023dark}.
	Note that the dark state is an eigenstate $\left\vert D \right\rangle $
	of the system that has no population in the atomic excited state and 
	 is annihilated by the atom-photon interaction Hamiltonian ${H}_{\text {int }}=\sum_{j=1}^{N}g_{j}\sigma _{-}a_{j}^{\dag }+g_{j}a_{j}\sigma _{+}$. Namely, the dark state satisfies the condition ${H}_{\text {int }}|D \rangle=0$.

	In general, the form of the coupling matrix is determined by both the Hamiltonian and the ordering for these upper and lower states. We point out that, for the multimode JC models under consideration, the coupling matrix can always be written as a row-echelon matrix with dimension $N_{u}\times N_{l}$ (with $N_{u}=C_{N+n-2}^{N-1}\text{ and } N_{l}=C_{N+n-1}^{N-1}$),
	\begin{equation}
	\small{	\mathbf{C}_{[N]}^{(n)}=\left( 
		\begin{array}{ccccc|c}
					\mathbf{M1}_{[N]}^{(n)} &\mathbf{\tilde{M}1}_{[N]}^{(n)} & \mathbf{0}& ...&\mathbf{0} & \mathbf{0}  \\ 
					\mathbf{0} & \mathbf{M2}_{[N]}^{(n)} &\mathbf{\tilde{M}2}_{[N]}^{(n)} & ...&\mathbf{0} & \mathbf{0} \\ 
					\mathbf{0} &	\mathbf{0} & \mathbf{M3}_{[N]}^{(n)}  & ...&\mathbf{0} & \mathbf{0} \\ 
					... & ...& ... & ...  & ...& ... \\ 
						\mathbf{0} & \mathbf{0} & \mathbf{0} &...& \mathbf{\tilde{M}(n-1)}_{[N]}^{(n)} &\mathbf{0}\\
					\mathbf{0} & \mathbf{0} & \mathbf{0} &...& \mathbf{Mn}_{[N]}^{(n)} &\mathbf{\tilde{M}n}_{[N]}^{(n)}
		\end{array}
		\right), }\label{CNn}
	\end{equation}%
	where $ \mathbf{0} $ denote zero matrices. 
	The matrix $\mathbf{C}_{[N]}^{(n)}$ is formed by block submatrices. The dimensions of the block submatrices $	\mathbf{M(s_{2}^{\prime}+1)}_{[N]}^{(n)} $ and $ \mathbf{\tilde{M}(s_{2}^{\prime}+1)}_{[N]}^{(n)} $ are $C_{N+s_{2}^{\prime}-2}^{N-2}\times C_{N+s_{2}^{\prime}-2}^{N-2}$ and $C_{N+s_{2}^{\prime}-2}^{N-2}\times C_{N+s_{2}^{\prime}-1}^{N-2}$, respectively. 
	In the following sections, we will present the expressions of these submatrices for the two-, three-, and four-mode JC models. As shown in Eq.~(\ref{CNn}), the coupling matrix can be divided into two parts [separated by vertical lines in Eq.~(\ref{CNn})], and the left part is an $N_{u} \times N_{u}$ upper triangular matrix, which helps us determine the rank and pivot columns of the coupling matrix. 
	
	In the coupling matrix, if a column has all zero elements, or can be reduced to a zero vector by elementary transformation, then the corresponding lower state is decoupled from all the upper states, and it becomes a dark state. Therefore, by using linear dependence to transform certain columns into $\mathbf{0}$ columns, we can obtain the form of the dark states. The existence of linear dependence between column vectors can be determined by calculating the column rank of the matrix (the number of the maximally linearly independent groups of the column vectors). If the number of columns is greater than the column rank, some columns can be represented as linear combinations of other columns, namely, these columns can be transformed into zero vectors. It means that we can find nonzero vectors $\mathbf{x}=(x_{1},x_{2},...,x_{N_{l}})^{T}$ to satisfy
	\begin{eqnarray}
			x_{1}\mathbf{C}_{[N]}^{(n)}{(:,1)}
		+x_{2}\mathbf{C}_{[N]}^{(n)} {(:,2)} +...
		+x_{N_{l}}\mathbf{C}_{[N]}^{(n)}{(:,N_{l})}  &=&\mathbf{0},\label{linear relationship}
	\end{eqnarray}%
	where \textquotedblleft$T$\textquotedblright\ denotes the matrix transpose and $\mathbf{C}_{[N]}^{(n)} {(:,s)}$ denotes the $s$th column of the matrix $\mathbf{C}_{[N]}^{(n)} $ with $s=1,2,...,N_{l}$. Equation~(\ref{linear relationship}) can be rewritten in a compact form as 
	\begin{eqnarray}
	\mathbf{C}_{[N]}^{(n)}\mathbf{x}=\mathbf{0},
    \end{eqnarray}%
	 then the set of all vectors $\mathbf{x}$ forms the null space of the matrix $\mathbf{C}_{[N]}^{(n)}$. Therefore, each solution $\mathbf{x}$ gives a set of superposition coefficients such that the linear superposition of some vectors is $\mathbf{0}$~\cite{Strang2009Introduction}. 
	
				\begin{figure*}[t!]
			\centering\includegraphics[width=0.86\textwidth]{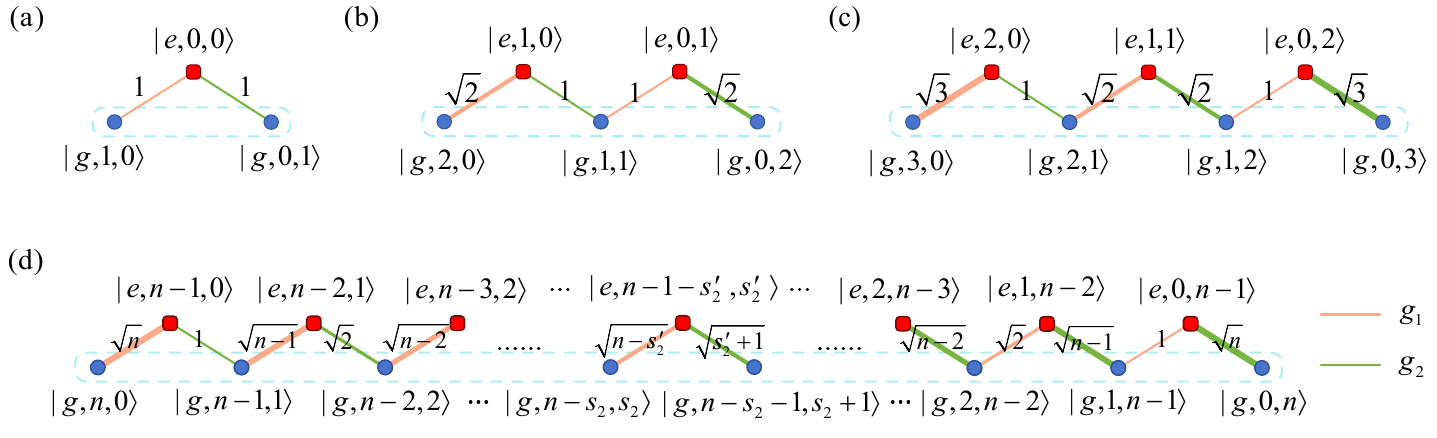}  
			\caption{The FSLs of the two-mode JC model restricted in the (a) single-, (b) double-, (c) triple-, and (d) $n$-excitation subspaces under the two-photon resonance condition $\Delta _{1}=\Delta _{2}=\Delta$. The red squares (blue circles) denote the upper (lower) states of the system in the given-excitation-number subspaces.}
			\label{twomode}
		\end{figure*}

	 For the considered multimode JC models, the coupling matrices take the form as 
	 	\begin{equation}
	 		\mathbf{C}_{[N]}^{(n)}=\left(
	 		\begin{array}{c|c}
	 			\underbrace{\begin{matrix}
	 					c_{11} &c_{12} & ... & c_{1N_{u}}   \\ 
	 					0 & c_{22} & ... & c_{2N_{u}} \\ 
	 					\vdots  & \vdots  & \ddots  & \vdots    \\ 
	 					0 & 0 & ... & c_{N_{u}N_{u}}\\
	 			\end{matrix}}_{\text{pivot columns}}
	 			&
	 			\underbrace{\begin{matrix}
	 					c_{1(N_{u}+1)}&...& c_{1N_{l}} \\ 
	 					c_{2(N_{u}+1)}&...& c_{2N_{l}} \\ 
	 					\vdots&\ddots&  \vdots \\ 
	 					c_{N_{u}(N_{u}+1)}&... & c_{N_{u}N_{l}} \\ 
	 			\end{matrix}}_{\text{free columns}}
	 		\end{array}	\right). \label{RE-CNn}
	 	\end{equation}%
	 We point out that the left part of the coupling matrix $\mathbf{C}_{[N]}^{(n)}$ is an upper triangular matrix.
	 	Then we can distinguish the pivot columns (which correspond to leading ones) from the free columns. According to the properties of  row-echelon matrices, we know that the rank of the coupling matrix is always equal to the number of rows. Therefore, there are $N_{u}$ linearly independent columns (pivot columns), and the remaining $ N_{l}-N_{u}$ free columns can be expressed as linear combinations of these linearly independent columns. According to Eq.~(\ref{RE-CNn}), we can obtain the new zero vectors that represent the dark states. For example, corresponding to the first column from the left in the free columns in Eq.~(\ref{RE-CNn}), we can obtain the relation 
	 	\begin{eqnarray}
	 	&&	\mathbf{C}_{[N]}^{(n)}{(:,N_{u}+1)}-x_{1}\mathbf{C}_{[N]}^{(n)}{(:,1)}
	 		-x_{2}\mathbf{C}_{[N]}^{(n)} {(:,2)} \notag\\
	 		&&-...
	 		-x_{N_{u}}\mathbf{C}_{[N]}^{(n)}{(:,N_{u})}  =\mathbf{0}	,\label{method}
	 	\end{eqnarray}%
	 	where we take a special case (otherwise the equation $\mathbf{C}_{[N]}^{(n)}\mathbf{x}=\mathbf{0}$ will have infinitely many solutions): setting the coefficient of the free column to 1, and then find the superposition coefficients of the pivot columns.
	 	By solving Eq.~(\ref{method}), the coefficients $x_{s=1-N_{u}}$ can be obtained. Then, one dark state can be obtained as a superposition of the basis states corresponding to these pivot columns and the superposition coefficient corresponding to the $s${th} pivot column is $x_{s}$.

	 	Using the same method, the dark states corresponding to all the free columns can be obtained. In general, there are $N_{l}-N_{u}$ dark states.
	 	In a realistic calculation, we can also directly determine the null space by  solving the equation $\mathbf{C}_{[N]}^{(n)}\mathbf{x}=\mathbf{0}$ using appropriate mathematical software. Based on the obtained vector $\mathbf{x}$, the dark states can be obtained. However, the dark states obtained in this method are not orthogonal and we can orthogonalize them with the  Gram-Schmidt orthogonalization. 
	 	In terms of the above method, we can establish the connection between the dark states and the null space of the coupling matrix.
	 	Note that the null space is a vector space defined as the set of all vectors $\mathbf{x}$ that satisfy the equation $\mathbf{C}_{[N]}^{(n)}\mathbf{x}=\mathbf{0}$. Although the solution is not unique, the set of these solutions (the null space) is unique. Therefore, the dark states obtained are not unique, but the dark-state subspace is unique.

	\section{Dark states in the two-mode JC model}\label{sec3}
	In this section, we study the dark states for the two-mode JC model, which is described by the Hamiltonian $\tilde{H}_{[2]}$ [Eq.~(\ref{Hamiltonian2}) for $N=2$]. Concretely, we study the dark states in the single-, double-, and triple-excitation subspaces. We also analyze the dark states in a general $n$-excitation subspace. In addition, we analyze the relationship between the dark states and dark modes, which are introduced based on the multimode JC Hamiltonians.

	\subsection{The dark states in the single-excitation subspace}
	In the single-excitation subspace, there are three basis states $%
	\{\left\vert e,0,0\right\rangle ,$ $\left\vert g,1,0\right\rangle ,$ $\left\vert
	g,0,1\right\rangle \}$ [Fig.~\ref{twomode}(a)]. According to the states of
	the atom, we confirm that there is one upper state $\left\vert
	e,0,0\right\rangle $ and two lower states $\left\vert g,1,0\right\rangle
	$ and $\left\vert g,0,1\right\rangle $. Namely, we can divide these three basis
	states into two subcomponents: the upper-state component $\left\vert
	e,0,0\right\rangle $ and the lower-state component $\{\left\vert
	g,1,0\right\rangle ,$ $\left\vert g,0,1\right\rangle\} $.
	
	We define the basis vectors for the basis states as $\left\vert e,0,0\right\rangle =(
	1,0,0) ^{T},$ $\left\vert g,1,0\right\rangle =( 0,1,0)
	^{T}, $ and $\left\vert g,0,1\right\rangle =( 0,0,1) ^{T}$. Then
	in the single-excitation subspace, the Hamiltonian $\tilde{%
		H}_{[2]}$ can be expressed as an arrowhead matrix 
	\begin{equation}
		\tilde{H}_{[2]}^{(1)}=\left( 
		\begin{array}{c|c}
			\mathbf{	U}_{[2]}^{(1)} & \mathbf{C}_{[2]}^{(1)} \\ \hline
			\left(\mathbf{C}_{[2]}^{(1) }\right)^{\dag } & \mathbf{L}_{[2]}^{(1)}%
		\end{array}
		\right) =\left( 
		\begin{array}{c|cc}
			0 & g_{1} & g_{2} \\ \hline
			g_{1} & -\Delta _{1} & 0 \\ 
			g_{2} & 0 & -\Delta _{2}%
		\end{array}
		\right) \label{H_21},
	\end{equation}%
	where $\mathbf{U}_{[2]}^{(1)}$ and $\mathbf{L}_{[2]}^{(1)}$ are, respectively, the submatrices related to the upper- and lower-state components in the single-excitation subspace, and  $\mathbf{C}_{[2]}^{(1)}$ is the coupling matrix describing the couplings
	between the two state components. Note that in Eq.~(\ref{H_21}) the superscript \textquotedblleft$(1)$\textquotedblright\ denotes the single-excitation subspace, and the subscript \textquotedblleft$[2]$\textquotedblright\ marks the two-mode JC model.
	
	According to the dark-mode theorems~\cite{huang2023dark}, the existence conditions
	of the dark states in the single-excitation subspace depend on the two
	detunings $\Delta _{1}$ and $\Delta _{2}$. In the two-photon resonance regime $%
	\Delta _{1}=\Delta _{2}=\Delta ,$ there is one dark state composed of two lower states, 
	\begin{equation}
		\left\vert D_{[2]}^{(1)}(1)\right\rangle =\frac{1}{\mathcal{N}_{[2]}^{(1)} }(
		g_{2}\left\vert g,1,0\right\rangle -g_{1}\left\vert g,0,1\right\rangle
		),
	\end{equation}%
	where we introduce the normalization constant $\mathcal{N}_{[2]}^{(1)} =(g_{1}^{2}+g_{2}^{2})^{1/2}$. The coupling configuration of the system in the single-excitation subspace is consistent with the $\Lambda$-type three-level system, and similarly the population could be coherently trapped in the two lower states under the two-photon resonance condition~\cite{ScullyQO1997}. Interestingly, the dark state $\left\vert D_{[2]}^{(1)}(1) \right\rangle$ can be expressed as a direct product of the atomic ground state $\left\vert g\right\rangle$ and the entangled state $(g_{2}\left\vert 1,0\right\rangle -g_{1}\left\vert 0,1\right\rangle)/{\mathcal{N}_{[2]}^{(1)} }$ of the two field modes. 
	
	If we introduce a mixing angle $\theta_{2} $ defined by $\tan \theta_{2}=g_{2}/g_{1}$, the dark state can be further expressed as
		\begin{equation}
			\left\vert D_{[2]}^{(1)}(1)\right\rangle =	\sin\theta_{2} \left\vert g,1,0\right\rangle -\cos\theta_{2}\left\vert g,0,1\right\rangle.	
		\end{equation}%
		It can be seen that by adiabatically tuning the mixing angle $\theta_{2} $ between $0$ and $\pi/2 $, it is possible to realize a complete and reversible state transfer between states $\left\vert g,0,1\right\rangle$ and $\left\vert g,1,0\right\rangle $.
	Therefore, we can use the STIRAP technique to realize a perfect single-photon transfer between the two field modes~\cite{STRIP2017}.
	
	\subsection{The dark states in the double-excitation subspace}
	In the double-excitation subspace, the basis states are $\{\left\vert
	e,1,0\right\rangle ,$ $\left\vert e,0,1\right\rangle ,$ $\left\vert
	g,2,0\right\rangle ,$ $\left\vert g,1,1\right\rangle,$ $\left\vert
	g,0,2\right\rangle \}$. There are two upper states $\{\left\vert
	e,1,0\right\rangle,$ $\left\vert e,0,1\right\rangle \}$ and three lower states 
	$\{\left\vert g,2,0\right\rangle ,$ $\left\vert g,1,1\right\rangle ,$ $\left\vert
	g,0,2 \right\rangle \}$ [Fig.~\ref{twomode}(b)]. By defining the basis
	vectors: $\left\vert e,1,0\right\rangle = ( 1,0,0,0,0) ^{T}$, $%
	\left\vert e,0,1\right\rangle =( 0,1,0,0,0) ^{T}$, $\left\vert
	g,2,0\right\rangle = ( 0,0,1,0,0) ^{T}$, $\left\vert
	g,1,1\right\rangle = ( 0,0,0,1,0) ^{T}, $ and $\left\vert
	g,0,2\right\rangle = ( 0,0,0,0,1) ^{T}$, the Hamiltonian $\tilde{H%
	}_{[2]}$ in the double-excitation subspace can be expressed as a thick
	arrowhead matrix 
	\begin{eqnarray}
		\tilde{H}_{[2]}^{(2)}&=&\left( 
		\begin{array}{c|c}
			\mathbf{U}_{[2]}^{(2)} & \mathbf{C}_{[2]}^{(2)} \\ \hline
			\left(\mathbf{C}_{[2]}^{(2) }\right)^{\dag } &\mathbf{L}_{[2]}^{(2)}%
		\end{array}
		\right)  \nonumber \\
		&=&\left( 
		\begin{array}{cc|ccc}
			-\Delta _{1} & 0 & \sqrt{2}g_{1} & g_{2} & 0 \\ 
			0 & -\Delta _{2} & 0 & g_{1} & \sqrt{2}g_{2} \\ \hline
			\sqrt{2}g_{1} & 0 & -2\Delta _{1} & 0 & 0 \\ 
			g_{2} & g_{1} & 0 & -\Delta _{1}-\Delta _{2} & 0 \\ 
			0 & \sqrt{2}g_{2} & 0 & 0 & -2\Delta _{2}%
		\end{array}
		\right) .  \label{H^2}
	\end{eqnarray}%
	Based on the dark-mode theorems, in the case of degenerate lower states, the
	existence and specific form of the dark states in the system can be obtained by
	analyzing the coupling matrix. In the case of $\Delta _{1}=\Delta _{2}=\Delta$, the matrix in Eq.~(\ref{H^2}) becomes 
	\begin{equation}
		\tilde{H}_{[2]}^{(2)}=\left( 
		\begin{array}{cc|ccc}
			-\Delta & 0 & \sqrt{2}g_{1} & g_{2} & 0 \\ 
			0 & -\Delta & 0 & g_{1} & \sqrt{2}g_{2} \\ \hline
			\sqrt{2}g_{1} & 0 & -2\Delta & 0 & 0 \\ 
			g_{2} & g_{1} & 0 & -2\Delta & 0 \\ 
			0 & \sqrt{2}g_{2} & 0 & 0 & -2\Delta%
		\end{array}
		\right) .
	\end{equation}%
	Mathematically, if we can transform a certain column of the coupling matrix $\mathbf{C}_{[2]}^{(2)}$ into a zero vector by elementary transformation, the corresponding superposed lower state is decoupled from all the upper states and becomes a dark state.
	Since the rank of the coupling matrix $\mathbf{C}_{[2]}^{(2)}$ is two, the coupling column vectors in $\mathbf{C}_{[2]}^{(2)}$ corresponding to the	states $\left\vert g,2,0\right\rangle ,$ $\left\vert g,1,1\right\rangle ,$ and $%
	\left\vert g,0,2\right\rangle $ are linearly dependent. Therefore, there is
	one dark state~\cite{huang2023dark}. Based on the relationship 
	\begin{equation}
		g_{2}^{2}\mathbf{C}_{[2]}^{(2)}{(:,1)} -\sqrt{2}g_{1}g_{2}\mathbf{C}_{[2]}^{(2)}{(:,2)} +g_{1}^{2}\mathbf{C}_{[2]}^{(2)}{(:,3)} =\mathbf{0},
	\end{equation}%
	we can obtain the superposition coefficients, and obtain the form of the dark state as
	\begin{equation}
		\left\vert D_{[2]}^{(2)}(1)\right\rangle =\frac{1}{\mathcal{N}_{[2]}^{(2)}}(
		g_{2}^{2}\left\vert g,2,0\right\rangle -\sqrt{2}g_{1}g_{2}\left\vert
		g,1,1\right\rangle +g_{1}^{2}\left\vert g,0,2\right\rangle ) , \label{D221}
	\end{equation}%
	where we introduce the normalization constant $\mathcal{N}_{[2]}^{(2)} =g_{1}^{2}+g_{2}^{2}	$. The coupling configuration of the two-mode JC model in the double-excitation subspace is consistent with the M-type five-level system. Similarly, the population can be coherently trapped in these three lower states under the two-photon resonance condition. 
	
	Using the mixing angle $\theta_{2} $, the dark state in Eq.~(\ref{D221}) can also be expressed as
		\begin{eqnarray}
			\left\vert D_{[2]}^{(2)}(1)\right\rangle &=&\sin^{2}\theta_{2}\left\vert g,2,0\right\rangle -\sqrt{2}\sin\theta_{2}\cos\theta_{2} \left\vert	g,1,1\right\rangle \notag \\
			&& +\cos^{2}\theta_{2}\left\vert g,0,2\right\rangle	.	
		\end{eqnarray}%
		By adiabatically tuning the mixing angle $\theta_{2} $ between $0$ and $\pi/2 $, we can realize a complete and reversible state transfer between the two states $\left\vert g,0,2\right\rangle$ and $\left\vert g,2,0\right\rangle $.	
	Similarly, with the STIRAP technique, we can realize a two-photon transfer between the two field modes by adiabatically modulating the coupling strengths $g_{1}$ and $g_{2}$.
	
	\subsection{The dark states in the triple-excitation subspace}
	In the triple-excitation subspace, the basis states are $%
	\{\left\vert e,2,0\right\rangle ,$ $\left\vert e,1,1\right\rangle ,$ $\left\vert
	e,0,2\right\rangle,$ $\left\vert g,3,0\right\rangle ,$ $\left\vert
	g,2,1\right\rangle $, $\left\vert g,1,2\right\rangle ,$ $\left\vert
	g,0,3\right\rangle \}$. There are three upper states $\{\left\vert
	e,2,0\right\rangle $, $\left\vert e,1,1\right\rangle,$ $\left\vert
	e,0,2\right\rangle \}$ and four lower states $\{\left\vert
	g,3,0\right\rangle ,$ $\left\vert g,2,1\right\rangle$, $\left\vert
	g,1,2\right\rangle,$ $ \left\vert g,0,3\right\rangle \}$ [Fig.~\ref{twomode}%
	(c)]. By defining the basis vectors: $\left\vert e,2,0\right\rangle =(
	1,0,0,0,0,0,0) ^{T}$, $\left\vert e,1,1\right\rangle =(
	0,1,0,0,0,0,0) ^{T}$, $\left\vert e,0,2\right\rangle =(
	0,0,1,0,0,0,0) ^{T}$, $\left\vert g,3,0\right\rangle =(
	0,0,0,1,0,0,0) ^{T}$, $\left\vert g,2,1\right\rangle =(
	0,0,0,0,1,0,0) ^{T}$, $\left\vert g,1,2\right\rangle =(
	0,0,0,0,0,1,0) ^{T},$ and $\left\vert g,0,3\right\rangle =(
	0,0,0,0,0,0,1) ^{T}$, the Hamiltonian $\tilde{H}_{[2]}$ in the
	triple-excitation subspace takes the form as 
	\begin{equation}
		\tilde{H}_{[2]}^{(3)}=\left( 
		\begin{array}{c|c}
			\mathbf{U}_{[2]}^{(3)} & \mathbf{C}_{[2]}^{(3)} \\ \hline
		\left(\mathbf{C}_{[2]}^{ (3)}\right)^{\dag } & \mathbf{L}_{[2]}^{(3)}%
		\end{array}
		\right) ,
	\end{equation}%
	where these submatrices are introduced as  
	\begin{subequations}
		\begin{align}
			\mathbf{U}_{[2]}^{(3)} =&\text{diag}(-2\Delta _{1},-\Delta _{1}-\Delta _{2},-2\Delta
			_{2}), \\
			\mathbf{L}_{[2]}^{(3)} =&\text{diag}(-3\Delta _{1},-2\Delta _{1}-\Delta _{2},-\Delta
			_{1}-2\Delta _{2},-3\Delta _{2}), \\
			\mathbf{C}_{[2]}^{(3)} =&\left( 
			\begin{array}{c|c|c|c}
				\sqrt{3}g_{1} & g_{2} & 0 & 0 \\ \hline
				0 & \sqrt{2}g_{1} & \sqrt{2}g_{2} & 0 \\ \hline
				0 & 0 & g_{1} & \sqrt{3}g_{2} 
			\end{array}
			\right)\notag \\
				=&\left( 
					\begin{array}{c|c|c|c}
						\mathbf{M1}_{[2]}^{(3)} & \mathbf{\tilde{M}1}_{[2]}^{(3)} & \mathbf{0} & \mathbf{0} \\ \hline
						\mathbf{0} & \mathbf{M2}_{[2]}^{(3)} &  \mathbf{\tilde{M}2}_{[2]}^{(3)} & \mathbf{0} \\ \hline
						\mathbf{0} & \mathbf{0} & \mathbf{M3}_{[2]}^{(3)} &  \mathbf{\tilde{M}3}_{[2]}^{(3)}%
					\end{array}
					\right).  \label{C23} 
		\end{align}%
	\end{subequations}
	
	The Hamiltonian $\tilde{H}_{[2]}^{(3)}$ is a thick arrowhead matrix and we can use the same
	method to analyze the dark states. The existence condition of the dark
	states is $\Delta _{1}=\Delta _{2}=\Delta $. Since the rank of the coupling
	matrix is three, the coupling column vectors in $\mathbf{C}_{[2]}^{(3)}$ corresponding to these four states $\left\vert g,3,0\right\rangle $, $\left\vert
	g,2,1\right\rangle $, $\left\vert g,1,2\right\rangle $, and $\left\vert
	g,0,3\right\rangle $ are linearly dependent, and there is one dark state
	consisting of these four lower states~\cite{huang2023dark}. Based on Eq.~(\ref{C23}), we can obtain the relationship
	\begin{eqnarray}
			&&g_{2}^{3}\mathbf{C}_{[2]}^{(3)}{(:,1)}
		-\sqrt{3}g_{2}^{2}g_{1}\mathbf{C}_{[2]}^{(3)} {(:,2)} 
		+\sqrt{3}g_{2}g_{1}^{2}\mathbf{C}_{[2]}^{(3)}{(:,3)} \notag \\
		&&-g_{1}^{3}\mathbf{C}_{[2]}^{(3)} {(:,4)}=\mathbf{0}.
	\end{eqnarray}%
	Then, we can obtain the form of the dark state
	\begin{eqnarray}
		\left\vert D_{[2]}^{(3)}(1)\right\rangle &=&\frac{1}{\mathcal{N}_{[2]}^{(3)}}
		(g_{2}^{3}\left\vert g,3,0\right\rangle -\sqrt{3}g_{2}^{2}g_{1}\left\vert
		g,2,1\right\rangle  \nonumber \\
		&&+\sqrt{3}g_{2}g_{1}^{2}\left\vert g,1,2\right\rangle -g_{1}^{3}\left\vert
		g,0,3\right\rangle ), \label{D231}
	\end{eqnarray}%
	where the normalization constant is introduced as $\mathcal{N}_{[2]}^{(3)}=
	(g_{1}^{2}+g_{2}^{2})^{3/2}.$ 
	
	In terms of the mixing angle $\theta_{2} $, the dark state in Eq.~(\ref{D231}) can be expressed as
		\begin{eqnarray}
			\small {\left\vert D_{[2]}^{(3)}(1)\right\rangle} &=&\small {\sin^{3}\theta_{2}\left\vert g,3,0\right\rangle -\sqrt{3}\sin^{2}\theta_{2}\cos\theta_{2}\left\vert	g,2,1\right\rangle } \nonumber \\
			&& \small{+\sqrt{3}\sin\theta_{2}\cos^{2}\theta_{2}\left\vert g,1,2\right\rangle -\cos^{3}\theta_{2}\left\vert	g,0,3\right\rangle .}	
		\end{eqnarray}%
		Similarly, we can realize a complete and reversible state transfer between the states $\left\vert g,0,3\right\rangle$ and $\left\vert g,3,0\right\rangle $ by adiabatically tuning the mixing angle $\theta_{2} \in [0,\pi/2 ]$.
	Using the STIRAP technique, the quantum transfer of three photons between the two modes can be realized by adiabatically modulating the coupling strengths $g_{1}$ and $g_{2}$.
	
	\subsection{The dark states in the $\boldsymbol{n}$-excitation subspace}
		
	For a general case, we analyze the dark states in the $n$-excitation subspace of the two-mode JC model [Fig.~\ref{twomode}(d)], in which
	there are $2n+1$ basis states, including $n$ upper states $\{\left\vert
	e,n-1,0\right\rangle ,$ $...,$ $\left\vert e,n-1-s_{2}^{\prime},s_{2}^{\prime}\right\rangle ,$ $...,$ $
	\left\vert e,0,n-1\right\rangle \} $ and $n+1$ lower states $\{\left\vert
	g,n,0\right\rangle ,$ $...,$ $\left\vert g,n-s_{2},s_{2}\right\rangle,$ $...,$ $\left\vert
	g,0,n\right\rangle \}$, where $s_{2}^{\prime}=0,1,2,...,n-1$ and $s_{2}=0,1,2,...,n$. Accordingly, we can define the basis vectors for these basis states as  
	\begin{subequations}
		\begin{align}
			\left\vert e,n-1,0\right\rangle & =(1,0,...,0,0)^{T}, \\
			\left\vert e,n-1-s_{2}^{\prime},s_{2}^{\prime}\right\rangle & =(0,0,...,1_{(s_{2}^{\prime}+1)},...,0,0)^{T}, \\
			\left\vert e,0,n-1\right\rangle & =(0,0,...,1_{(n)},...,0,0)^{T}, \\
			\left\vert g,n,0\right\rangle & =(0,0,...,1_{(n+1)},...,0,0)^{T}, \\
			\left\vert g,n-s_{2},s_{2}\right\rangle & =(0,0,...,1_{(n+s_{2}+1)},...,0,0)^{T}, \\
			\left\vert g,0,n\right\rangle & =(0,0,...,0,1)^{T},
		\end{align}%
	\end{subequations}
	where the subscript of the element 1 is introduced to indicate the position of the element 1. The Hamiltonian $\tilde{H}_{[2]}$ in the $n$-excitation subspace can be
	expressed as 
	\begin{equation}
		\tilde{H}_{[2]}^{(n)}=\left( 
		\begin{array}{c|c}
			( \mathbf{U}_{[2]}^{(n)}) _{n\times n} & ( \mathbf{C}_{[2]}^{(n)})
			_{n\times (n+1)} \\ \hline
			 \left(\mathbf{C}_{[2]}^{ (n)}\right)^{\dag } _{(n+1)\times n} & (
			\mathbf{L}_{[2]}^{(n)}) _{(n+1)\times (n+1)}%
		\end{array}
		\right) ,\label{H2n}
	\end{equation}%
	where we add the subscripts to denote the dimension of these submatrices. In Eq.~(\ref{H2n}), these submatrices are defined by  
	\begin{subequations}
		\begin{align}
			\mathbf{U}_{[2]}^{(n)} =&\text{diag}(-(n-1)\Delta _{1},...,-(n-1-s_{2}^{\prime})\Delta
			_{1}-s_{2}^{\prime}\Delta _{2}, \notag \\
			& ...,-(n-1)\Delta _{2}), \\
			\mathbf{L}_{[2]}^{(n)}=&\text{diag}(-n\Delta _{1},...,-(n-s_{2})\Delta _{1}-s_{2}\Delta
			_{2},...,-n\Delta _{2}), \\
			\mathbf{C}_{[2]}^{(n)}=&\small{{\left(
					\begin{array}{c|c|c|c|c|c}
						\sqrt{n}g_{1} & g_{2} & 0 & 0 & 0 & 0 \\ \hline
						0 & \cdots & \cdots & 0 & 0 & 0 \\ \hline
						0 & 0 &  \sqrt{n-s_{2}^{\prime}}g_{1} & \sqrt{s_{2}^{\prime}+1}g_{2} & 0 & 0 \\ \hline
						0 & 0 & 0 & \cdots & \cdots & 0 \\ \hline
						0 & 0 & 0 & 0 & g_{1} & \sqrt{n}g_{2}%
					\end{array}
					\right) }} \notag \\
			=&{\small{\left(
					\begin{array}{c|c|c|c|c|c}
						\mathbf{M1}_{[2]}^{(n)} & \mathbf{\tilde{M}1}_{[2]}^{(n)} & 0 & 0 & 0 & 0 \\ \hline
						0 & \cdots & \cdots & 0 & 0 & 0 \\ \hline
						0 & 0 & \mathbf{M(s_{2}^{\prime}+1)}_{[2]}^{(n)} & \mathbf{\tilde{M}(s_{2}^{\prime}+1)}_{[2]}^{(n)} & 0 & 0 \\ \hline
						0 & 0 & 0 & \cdots & \cdots & 0 \\ \hline
						0 & 0 & 0 & 0 & \mathbf{Mn}_{[2]}^{(n)} & \mathbf{\tilde{M}n}_{[2]}^{(n)}%
					\end{array}
					\right)} }.
		\end{align}%
	\end{subequations}
	
	In the two-photon resonance regime $\Delta _{1}=\Delta_{2}=\Delta $, these $n+1$ lower states are degenerate. Since the rank of the coupling matrix $\mathbf{C}_{[2]}^{(n)}$
	is $n$, these $n+1$ column vectors in the coupling matrix are linearly dependent. Next, we find the linear dependence among these vectors. We assume that these $n+1$ linearly dependent vectors satisfy
	\begin{equation}
		\sum_{s_{2}=0}^{n}x_{s_{2}} \mathbf{C}_{[2]}^{(n)}({:,s_{2}+1})=\mathbf{0}, \label{method1}
	\end{equation}%
	where $x_{s_{2}}$ are the coefficients to be determined. 
	We point out that the basis state corresponding to the only free column in $\mathbf{C}_{[2]}^{(n)}$ is $\left\vert g,0,n\right\rangle$ (with $s_{2}=n$). By solving Eq.~(\ref{method1}), we can obtain the dark state as 	
	\begin{equation}
	\small {	\left\vert D_{[2]}^{(n)}(1)\right\rangle	=\frac{1}{\mathcal{N}_{[2]}^{(n)}}\sum_{k_{2}=0}^{n}\sqrt{\frac{n!}{%
				(n-k_{2})!k_{2}! }}( -g_{1})^{n-k_{2}}g_{2} ^{k_{2}}\left\vert
		g,k_{2},n-k_{2}\right\rangle,}  \label{Dn'}
	\end{equation}%
	where we introduce the normalization constant $\mathcal{N}_{[2]}^{(n)}=
	(g_{1}^{2}+g_{2}^{2})^{n/2}$. Equation~(\ref{Dn'}) presents a general formula for the dark
	state in the two-mode JC model. For any $n$ excitations, there is always only one dark state, which is composed of the lower states $\left\vert g,k_{2},n-k_{2}\right\rangle$. Therefore, the dark state does not contain the atomic excited state and it is immune to spontaneous emission~\cite{ScullyQO1997}.

	In terms of the mixing angle $\theta_{2} $, the dark state in Eq.~(\ref{Dn'}) can be expressed as
	\begin{eqnarray}
		\left\vert D_{[2]}^{(n)}(1)\right\rangle &=&\sum_{k_{2}=0}^{n}\sqrt{\frac{n!}{(n-k_{2})!k_{2}!}}
		(-1)^{n-k_{2}}\cos ^{n-k_{2}}(\theta_{2})\notag \\
		&&\times \sin^{k_{2}}(\theta_{2})\left\vert g,k_{2},n-k_{2}\right\rangle.
		\label{Dn_theta}
	\end{eqnarray}%
	It can be seen from Eq.~(\ref{Dn_theta}) that, by adiabatically tuning the mixing angle $\theta_{2} $ from $0$ to $\pi/2 $, it is possible to realize complete and reversible state transfer from state $\left\vert g,0,n\right\rangle $ to $\left\vert g,n,0\right\rangle$. This means the transfer of $n$ excitations from mode $a_{2}$ to mode $a_{1}$.
	
	It is worth noting that the FSL corresponding to the two-mode JC model takes the form of a one-dimensional Su-Schrieffer-Heeger (SSH)-like lattice. Based on the definition of the dark states, which satisfies ${H}_{\text {int }}|D \rangle=0$ and has no population in the atomic excited state, we can see that the dark state obtained here is related to the topological zero-energy state in the system~\cite{TP:NC2021}.
	
		\begin{figure*}[t]
		\centering\includegraphics[width=0.9\textwidth]{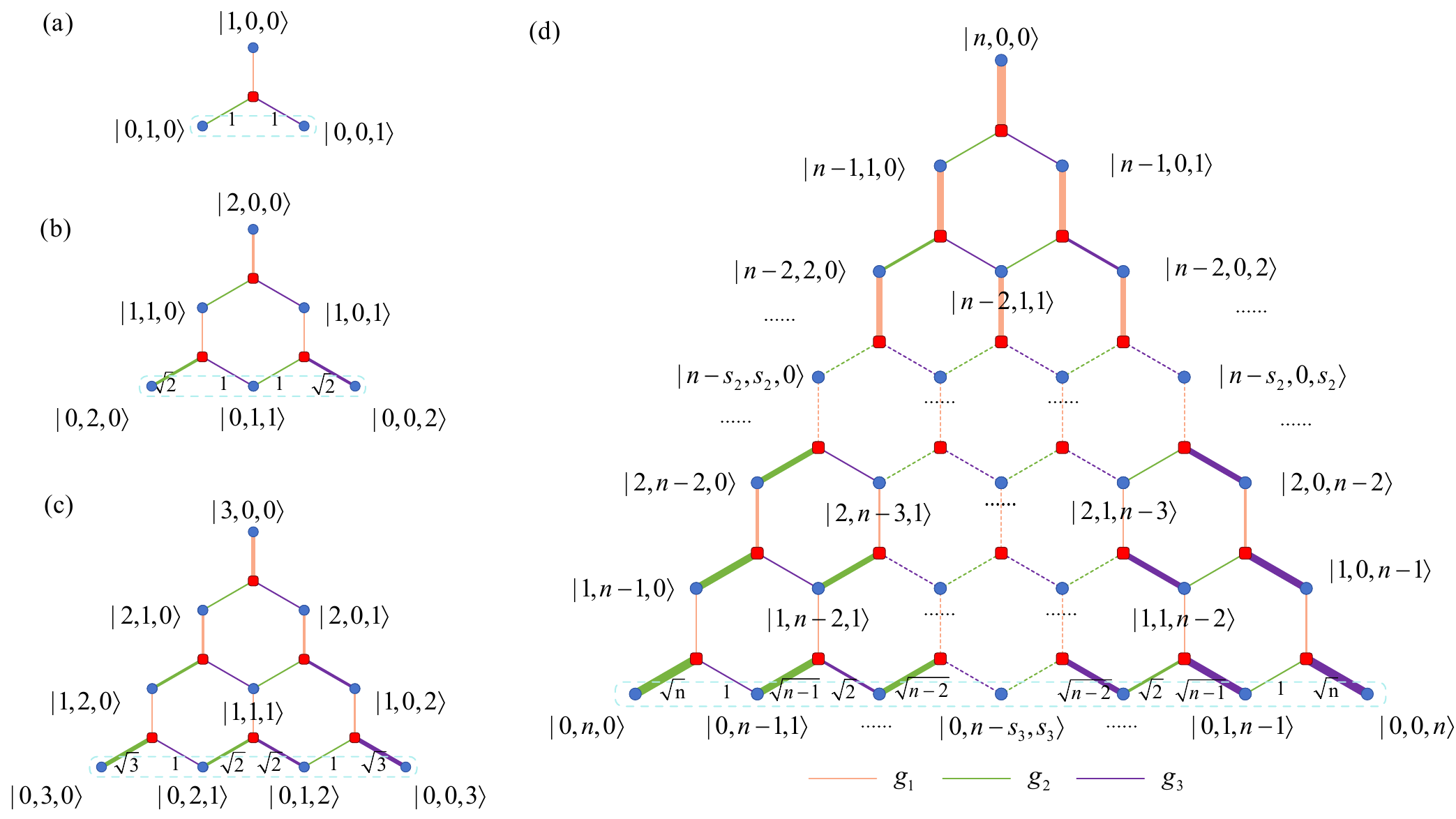}  
		\caption{The FSLs of the three-mode JC model restricted in the (a) single-, (b) double-, (c) triple-, and (d) $n$-excitation subspaces under the resonance condition $\Delta _{1}=\Delta _{2}=\Delta _{3}=\Delta $. The red squares (blue circles) denote the upper (lower) states of the system in the given-excitation-number subspaces. For clarity, we only label the lower states in these FSLs.}
		\label{threemode}
	\end{figure*}

	\subsection{The relationship between the dark states and dark modes}
	We point out that the dark-state effect can also be investigated by analyzing the dark modes in the multimode JC models.
	For the two-mode JC model, we can show that when the two field modes are degenerate, there exists a dark-mode effect. Concretely, for the case $\omega _{1}=\omega _{2}=\omega _{c}$, we can introduce the bright mode and dark mode for the two-mode JC model. To this end, we introduce~\cite{huang2023dark}
	\begin{equation}
		a_{2+}=\frac{1}{\mathcal{N}_{[2]}^{(1)} }(g_{1}a_{1}+g_{2}a_{2}) ,\hspace{0.3cm}
		a_{2-}=\frac{ 1}{\mathcal{N}_{[2]}^{(1)} }(g_{2}a_{1}-g_{1}a_{2}),\label{a+a-}
	\end{equation}%
	which satisfy the relations $[a_{2+},a_{2+}^{\dag}]=1$, $[a_{2-},a_{2-}^{\dag}]=1$, and $[a_{2+},a_{2-}^{\dag}]=0$. In the representation of these new modes, the Hamiltonian of the two-mode JC model can be expressed as
	\begin{equation}
		H_{[2]}=\frac{\omega _{0}}{2}\sigma _{z}+\omega _{c}( a_{2+}^{\dag
		}a_{2+}+a_{2-}^{\dag }a_{2-}) +\mathcal{N}_{[2]}^{(1)} ( \sigma
		_{-}a_{2+}^{\dag }+a_{2+}\sigma _{+}) ,  \label{H_2}
	\end{equation}%
	with the relationship  $a_{2+}^{\dag}a_{2+}+a_{2-}^{\dag}a_{2-}=a_{1}^{\dag}a_{1}+a_{2}^{\dag}a_{2}$ in the case of $\omega _{1}=\omega _{2}=\omega _{c}$.
	We can see from Eq.~(\ref{H_2}) that the mode $a_{2-}$ is a dark mode, because it decouples from both the mode $a_{2+}$ and the atom. We can prove that only when the degeneracy condition is satisfied, the mode $a_{2-}$ can be decoupled from both the mode $a_{2+}$ and the atom~\cite{huang2023dark}.
	
	At the same time, we can obtain the relationship between the vacuum states in the $(a_{2+},a_{2-})$- and $(a_{1},a_{2})$-mode representations: $\left\vert
	0\right\rangle _{a_{2+}}\left\vert 0\right\rangle _{a_{2-}}=\left\vert
	0\right\rangle _{a_{1}}\left\vert 0\right\rangle _{a_{2}}$. Further, in terms of the relations in Eq.~(\ref{a+a-}), we can obtain the relationship between the basis states in the two representations as
	\begin{eqnarray}
	&&\left\vert n-m_{2}\right\rangle _{a_{2+}}\left\vert m_{2}\right\rangle _{a_{2-}} \nonumber \\
		&= &
	\frac{a_{2+}^{\dag(n-m_{2})} a_{2-}^{{\dag}m_{2}}}{\sqrt{(n-m_{2})!m_{2}!}}\left\vert
		0\right\rangle _{a_{2+}}\left\vert 0\right\rangle _{a_{2-}} \nonumber \\
		& =&	\frac {\sqrt{(n-m_{2}) !m_{2}!}}{ \mathcal{N}_{[2]}^{(n)} } \sum
			_{k_{2}=0}^{n}C_{k_{2},m_{2}}^{n}\left\vert k_{2} \right\rangle _{a_{1}}\left\vert
			n-k_{2}\right\rangle _{a_{2}},\label{Cmn,N}
	\end{eqnarray}%
	where $m_{2}=0,1,2,...,n$, $k_{2}=0,1,2,...,n$, and the coefficients $C_{k_{2},m_{2}}^{n}$ are given by%
	\begin{equation}
		\small {C_{k_{2},m_{2}}^{n}=\sum_{q=\max ( 0,m_{2}-n+k_{2})}^{\min ( k_{2},m_{2})}\frac{\sqrt{(n-k_{2})! k_{2} !}(
			-1) ^{m_{2}-q}g_{1}^{m_{2}-2q+k_{2}}g_{2}^{n-k_{2}-m_{2}+2q}}{( k_{2}-q) !(
			n-k_{2}-m_{2}+q) !q!( m_{2}-q) !}.}
	\end{equation}%

	Next, we can infer the form of the dark state in the $n$-excitation subspace. \textit{For the $n$-excitation case, we know that all these excitations should be stored in the dark mode $a_{2-}$. This is because, if there are excitations in the bright mode $a_{2+}$, then these excitations will be transferred to the atomic excited state $\left\vert e\right\rangle $ via the coupling term $\mathcal{N}_{[2]}^{(1)} ( \sigma
		_{-}a_{2+}^{\dag }+a_{2+}\sigma _{+}) $. However, we know that the dark state only contains the ground-state component of the atom. Based on the above analysis, we know that the dark state in the $n$-excitation subspace should take the form of
		$\left\vert g\right\rangle  \left\vert 0\right\rangle _{a_{2+}}\left\vert
		n\right\rangle _{a_{2-}} $} [\textit{with $m_{2}=n$ in Eq.~(\ref{Cmn,N})}]. In terms of Eq.~(\ref{Cmn,N}), we can write out the expression of the dark state as
	\begin{eqnarray}
	\left\vert D_{[2]}^{(n)}(1)\right\rangle&=&	\left\vert g\right\rangle \left\vert 0\right\rangle _{a_{2+}}\left\vert		n\right\rangle _{a_{2-}} =\frac{a_{2-}^{\dag n}}{\sqrt{n!}}\left\vert g\right\rangle \left\vert 0\right\rangle _{a_{2+}}\left\vert 0\right\rangle _{a_{2-}}.
	\end{eqnarray}%
	 This result is the same as the previously obtained result given in Eq.~(\ref{Dn'}).

	 It is worth noting that both the dark-mode method and the arrowhead-matrix method have their individual advantages in analytical approaches and applicability. 
	 The key point of the dark-mode method is to construct the dark mode decoupled from the target part, which is particularly efficient in systems with explicit symmetries or well-structured systems. The arrowhead-matrix method does not depend on the symmetry of the system and has strong universality, but it requires dealing with complex matrix operations.
	 In fact, the dark-mode method is a "top-down" method, while the arrowhead-matrix method is a "bottom-up" method. It can be seen that the dark-mode method is simpler in operation than the arrowhead-matrix method. But if one aims to know the detailed transitions among these basis states for coding quantum information, the arrowhead-matrix method has more advantages in this context, which is also the necessity of introducing the arrowhead-matrix method in this paper.

	\section{Dark states in the three-mode JC model}\label{sec4}
	
	Next we consider the case of the three-mode JC model depicted by the Hamiltonian $\tilde{H}_{[3]}$ [Eq.~(\ref{Hamiltonian2}) for $N=3$]. Similarly, we first analyze the dark states in the single-, double-, and triple-excitation subspaces, and then analyze the coupling matrix to find the general formula for the dark states in the $n$-excitation subspace.
	
	\subsection{The dark states in the single-excitation subspace}
	In the single-excitation subspace [Fig.~\ref{threemode}(a)], the basis
	states are given by $\{\left\vert e,0,0,0\right\rangle ,$ $\left\vert
	g,1,0,0\right\rangle ,$ $\left\vert g,0,1,0\right\rangle ,$ $\left\vert
	g,0,0,1\right\rangle \}$, and there is one upper state $\left\vert
	e,0,0\right\rangle $ and three lower states $\{\left\vert
	g,1,0,0\right\rangle ,$ $\left\vert g,0,1,0\right\rangle $, $\left\vert
	g,0,0,1\right\rangle \}$. We define the basis vectors: $\left\vert
	e,0,0,0\right\rangle =( 1,0,0,0) ^{T}$, $\left\vert
	g,1,0,0\right\rangle =( 0,1,0,0) ^{T}$, $\left\vert
	g,0,1,0\right\rangle =( 0,0,1,0) ^{T},$ and $\left\vert
	g,0,0,1\right\rangle =( 0,0,0,1) ^{T}$, then the Hamiltonian $%
	\tilde{H}_{[3]}$ in the single-excitation subspace can be expressed as 
	\begin{equation}
		\tilde{H}_{[3]}^{(1)}=\left( 
		\begin{array}{c|ccc}
			0 & g_{1} & g_{2} & g_{3} \\ \hline
			g_{1} & -\Delta _{1} & 0 & 0 \\ 
			g_{2} & 0 & -\Delta _{2} & 0 \\ 
			g_{3} & 0 & 0 & -\Delta _{3}%
		\end{array}%
		\right) .
	\end{equation}%
	Here, the coupling matrix is a row vector, and there are two sets of linear dependence with the first column denoted as the pivot column. In the case of  $\Delta _{1}=\Delta _{2}=\Delta _{3}=\Delta $, we can obtain the following two dark states:
	\begin{subequations}
		\begin{align}
			\left\vert D_{[3]}^{(1)}(1) \right\rangle & =\frac{1}{\sqrt{g_{2}^{2}+g_{1}^{2}}}(
			g_{2}\left\vert g,1,0,0\right\rangle -g_{1}\left\vert g,0,1,0\right\rangle
			) , \\
			\left\vert D_{[3]}^{(1)}(2)\right\rangle & =\frac{1}{\sqrt{g_{3}^{2}+g_{1}^{2}}}(
			g_{3}\left\vert g,1,0,0\right\rangle -g_{1}\left\vert g,0,0,1\right\rangle
			).
		\end{align} \label{D_3_1}%
	\end{subequations}
	
	Note that the dark states are not unique because the linear dependence is not unique. These two dark states can be orthogonalized by using the Gram-Schmidt orthogonalization. Then we obtain the normalized orthogonal dark states as
	\begin{subequations}
		\begin{align}
			\left\vert \tilde{D}_{[3]}^{(1)}(1)\right\rangle =&\frac{1}{\mathcal{N}_{[2]}^{(1)} }(g_{2}\left\vert
			g,1,0,0\right\rangle -g_{1}\left\vert g,0,1,0\right\rangle ),   \\
			\left\vert \tilde{D}_{[3]}^{(1)}(2)\right\rangle  =&\frac{1}{\mathcal{N}_{[3]}^{(1)} \mathcal{N}_{[2]}^{(1)} }%
			(g_{1}g_{3}\left\vert g,1,0,0\right\rangle+g_{2}g_{3}\left\vert
			g,0,1,0\right\rangle \notag\\
			&	-\mathcal{N}_{[2]}^{(2)}\left\vert g,0,0,1\right\rangle ),
		\end{align} \label{D1D2}%
	\end{subequations}
	where we introduce the normalization constant $\mathcal{N}_{[3]}^{(1)} =(g_{1}^{2}+g_{2}^{2}+g_{3}^{2})^{1/2}$. Notice that the two linearly independent dark states in Eqs.~(\ref{D_3_1}) span a two-dimensional subspace of dark states. After orthogonalization, the two states in Eqs.~(\ref{D1D2}) can be used as a set of orthogonal dark states for this subspace, and any unitary transformation of them can represent a new set of orthogonal dark states. Therefore, the orthogonal dark states we obtain are not unique, but the dark-state subspace is unique. We emphasize that for degenerate dark states, the ordinary STIRAP techniques cannot be used to realize quantum state transfer. This is because the parameter conditions for the quantum adiabatic evolution do not satisfy the degenerate states. Nevertheless, the schemes for implementing STIRAP in the degenerate subspaces have been proposed~\cite{STIRAPiDEG2001,STIRAPiDEG2008,STIRAPiDEG2013}.

	\subsection{The dark states in the double-excitation subspace}
	In the double-excitation subspace [Fig.~\ref{threemode}(b)], the basis 
	states are $\{\left\vert e,1,0,0\right\rangle ,$ $\left\vert 
	e,0,1,0\right\rangle ,$ $\left\vert e,0,0,1\right\rangle ,$ $\left\vert 
	g,2,0,0\right\rangle ,$ $\left\vert g,1,1,0\right\rangle $, $\left\vert 
	g,1,0,1\right\rangle $, $\left\vert g,0,2,0\right\rangle $, $\left\vert 
	g,0,1,1\right\rangle $, $\left\vert g,0,0,2\right\rangle \}$, and there are three
	upper states associated with the atomic excited state $\left\vert e\right\rangle$ and six lower states associated with the atomic ground state $\left\vert g\right\rangle$. We define the basis vectors: $\left\vert 
	e,1,0,0\right\rangle =(1,0,0,0,0,0,0,0,0)^{T}$, $\left\vert 
	e,0,1,0\right\rangle =(0,1,0,0,0,0,0,0,0)^{T}$, $\left\vert 
	e,0,0,1\right\rangle =(0,0,1,0,0,0,0,0,0)^{T}$, $\left\vert 
	g,2,0,0\right\rangle =(0,0,0,1,0,0,0,0,0)^{T}$, $\left\vert 
	g,1,1,0\right\rangle =(0,0,0,0,1,0,0,0,0)^{T}$, $\left\vert 
	g,1,0,1\right\rangle =(0,0,0,0,0,1,0,0,0)^{T}$, $\left\vert 
	g,0,2,0\right\rangle =(0,0,0,0,0,0,1,0,0)^{T}$, $\left\vert 
	g,0,1,1\right\rangle =(0,0,0,0,0,0,0,1,0)^{T}$, and $\left\vert 
	g,0,0,2\right\rangle =(0,0,0,0,0,0,0,0,1)^{T}$. Then the Hamiltonian $\tilde{H}
	_{[3]}$ in the double-excitation subspace can be expressed as 
	\begin{equation}
		\tilde{H}_{[3]}^{(2)}=\left( 
		\begin{array}{c|c}
			\mathbf{U}_{[3]}^{(2)} & \mathbf{C}_{[3]}^{(2)} \\ \hline
			\left(\mathbf{C}_{[3]}^{ (2)}\right)^{\dag } & \mathbf{L}_{[3]}^{(2)}%
		\end{array}
		\right) ,
	\end{equation}%
	where these submatrices are defined as  
	\begin{subequations}
		\begin{align}
			\mathbf{U}_{[3]}^{(2)} =&\text{diag}(-\Delta _{1},-\Delta _{2},-\Delta _{3}), \\
			\mathbf{L}_{[3]}^{(2)}=& \text{diag}(-2\Delta _{1},-\Delta _{1}-\Delta _{2},-\Delta
			_{1}-\Delta _{3}, -2\Delta _{2},  \nonumber \\
			&  -\Delta _{2}-\Delta _{3},-2\Delta _{3}), \\
			\mathbf{C}_{[3]}^{(2)} =&\left( 
			\begin{array}{c|cc|ccc}
				\sqrt{2}g_{1} & g_{2} & g_{3} & 0 & 0 & 0 \\ \hline
				0 & g_{1} & 0 & \sqrt{2}g_{2} & g_{3} & 0 \\ 
				0 & 0 & g_{1} & 0 & g_{2} & \sqrt{2}g_{3}%
			\end{array}
			\right) \notag\\
			=&\left( 
			\begin{array}{c|c|c}
				\mathbf{M1}_{[3]}^{(2)} & 	\mathbf{\tilde{M}1}_{[3]}^{ (2)} & \mathbf{0} \\ \hline
				\mathbf{0} & 	\mathbf{M2}_{[3]}^{(2)} &	\mathbf{\tilde{M}2}_{[3]}^{ (2)}%
			\end{array}
			\right) .\label{C_3_2}
		\end{align}%
	\end{subequations}
	Notice that $ \mathbf{M1}_{[3]}^{(2)}$, $ \mathbf{\tilde{M}1}_{[3]}^{ (2)}$, $ \mathbf{M2}_{[3]}^{(2)}$, $ \mathbf{\tilde{M}2}_{[3]}^{ (2)}$, and $\mathbf{0} $ are matrices. 
	
	The rank of the coupling matrix $\mathbf{C}_{[3]}^{(2)}$ is three, and $	\mathbf{L}_{[3]}^{(2)}$ is a six-dimensional degenerate subspace when $\Delta_{1}=\Delta _{2}=\Delta _{3}=\Delta $. In this case, the first three columns are pivot columns, and the last three are free columns.
	Therefore, we can obtain three groups of linearly dependent vectors, and the dark states can be written as
	\begin{subequations}
		\begin{align}
		 {	\left\vert D_{[3]}^{(2)}(1)\right\rangle} =& {g_{2}^{2}\left\vert g,2,0,0\right\rangle -
			\sqrt{2}g_{1}g_{2}\left\vert g,1,1,0\right\rangle +g_{1}^{2}\left\vert
			g,0,2,0\right\rangle }, \\
			 {\left\vert D_{[3]}^{(2)}(2)\right\rangle } =& {\sqrt{2}g_{2}g_{3}\left\vert
			g,2,0,0\right\rangle -g_{1}g_{3}\left\vert g,1,1,0\right\rangle
			-g_{1}g_{2}\left\vert g,1,0,1\right\rangle}\notag \\
			&+ {g_{1}^{2}\left\vert
			g,0,1,1\right\rangle }, \\
			 {\left\vert D_{[3]}^{(2)}(3)\right\rangle } =& {g_{3}^{2}\left\vert g,2,0,0\right\rangle -%
			\sqrt{2}g_{3}g_{1}\left\vert g,1,0,1\right\rangle +g_{1}^{2}\left\vert
			g,0,0,2\right\rangle }.
		\end{align}\label{D_3_2}%
	\end{subequations}
	
	The dark states shown in Eqs.~(\ref{D_3_2}) are not normalized. In principle, we can orthogonalize and normalize these dark states by using the Gram-Schmidt orthogonalization. Based on Eqs.~(\ref{D_3_2}), we obtain the normalized orthogonal dark states
	\begin{subequations}
		\begin{align}
	 {	\left\vert \tilde{D}_{[3]}^{(2)}(1)\right\rangle} =&  {\frac{1}{\mathcal{N}%
			_{[2]}^{(2)}}(g_{2}^{2}\left\vert g,2,0,0\right\rangle -\sqrt{2}%
		g_{1}g_{2}\left\vert g,1,1,0\right\rangle  } \notag \\
		&  {+g_{1}^{2}\left\vert g,0,2,0\right\rangle )}, \\
			 {\left\vert \tilde{D}_{[3]}^{(2)}(2)\right\rangle} =&  {\frac{1}{\mathcal{N}%
			_{[2]}^{(2)}\mathcal{N}_{[3]}^{(1)}}(\sqrt{2}g_{1}g_{2}g_{3}\left\vert
		g,2,0,0\right\rangle  } \notag \\
		&  {+g_{3}(g_{2}^{2}-g_{1}^{2})\left\vert g,1,1,0\right\rangle -\mathcal{N}%
		_{[2]}^{(2)}g_{2}\left\vert g,1,0,1\right\rangle}  \notag \\
		& { +\sqrt{2}g_{1}g_{2}g_{3}\left\vert g,0,2,0\right\rangle +\mathcal{N}%
		_{[2]}^{(2)}g_{1}\left\vert g,0,1,1\right\rangle )}, \\
		 {\left\vert \tilde{D}_{[3]}^{(2)}(3)\right\rangle} =&  {\frac{1}{\mathcal{N}%
			_{[2]}^{(2)}\mathcal{N}_{[3]}^{(2)}}(g_{1}^{2}g_{3}^{2}\left\vert
		g,2,0,0\right\rangle  +%
		\mathcal{N}_{[2]}^{(4)}\left\vert g,0,0,2\right\rangle } \notag \\
		&  {+\sqrt{2}g_{1}g_{2}g_{3}^{2}\left\vert g,1,1,0\right\rangle-\sqrt{2}\mathcal{N}_{[2]}^{(2)}g_{1}g_{3}\left\vert g,1,0,1\right\rangle  } \notag \\
		& { +g_{2}^{2}g_{3}^{2}\left\vert g,0,2,0\right\rangle-\sqrt{2}\mathcal{N}_{[2]}^{(2)}g_{2}g_{3}\left\vert g,0,1,1\right\rangle  	  )},
		\end{align}\label{D1D2D3}%
	\end{subequations} %
	where we introduce the normalization constant $\mathcal{N}_{[3]}^{(2)} =g_{1}^{2}+g_{2}^{2}+g_{3}^{2}$. 

	\subsection{The dark states in the triple-excitation subspace}
	In the triple-excitation subspace [Fig.~\ref{threemode}(c)], the basis states are $\{\left\vert e,2,0,0\right\rangle $, $\left\vert e,1,1,0 \right\rangle  $, $\left\vert e,1,0,1\right\rangle $, $\left\vert e,0,2,0 \right\rangle $, $	\left\vert e,0,1,1\right\rangle $, $\left\vert e,0,0,2\right\rangle $, $\left\vert g,3,0,0\right\rangle $, $\left\vert g,2,1,0\right\rangle $, $	\left\vert g,2,0,1\right\rangle $, $\left\vert g,1,2,0\right\rangle $, $\left\vert g,1,1,1\right\rangle $, $\left\vert g,1,0,2\right\rangle $, $\left\vert g,0,3,0\right\rangle $, $\left\vert g,0,2,1\right\rangle $, $\left\vert g,0,1,2\right\rangle $, $\left\vert g,0,0,3\right\rangle \}$, and there are six upper states associated with the atomic excited state $\left\vert e\right\rangle$ and ten lower states associated with the atomic ground state $\left\vert g\right\rangle$. We define the basis vectors
	corresponding to these basis states as $( 1,0,0,0,...,0) ^{T}$, $	( 0,1,0,0,...,0) ^{T}$, ..., and  $( 0,0,0,...,0,1) ^{T}$, then the Hamiltonian $\tilde{H}_{[3]}$ in the triple-excitation subspace takes the form 
	\begin{equation}
		\tilde{H}_{[3]}^{(3)}=\left( 
		\begin{array}{c|c}
			\mathbf{U}_{[3]}^{(3)} & \mathbf{C}_{[3]}^{(3)} \\ \hline
			\left(\mathbf{C}_{[3]}^{ (3)}\right)^{\dag } & \mathbf{L}_{[3]}^{(3)}%
		\end{array}
		\right) ,
	\end{equation}%
	where these submatrices are introduced as  
	\begin{subequations}
		\begin{align}
			\mathbf{U}_{[3]}^{(3)} =&\text{diag}(-2\Delta _{1},-\Delta _{1}-\Delta _{2},-\Delta
			_{1}-\Delta _{3}, -2\Delta _{2},  -\Delta _{2}-\Delta _{3}, \nonumber \\
			& -2\Delta _{3}), \\
			\mathbf{L}_{[3]}^{(3)} =&\text{diag}(-3\Delta _{1},-2\Delta _{1}-\Delta _{2},-2\Delta
			_{1}-\Delta _{3}, -\Delta_{1}-2\Delta _{2}, -\Delta _{1} \nonumber \\
			&-\Delta _{2} -\Delta _{3}, -\Delta _{1}-2\Delta _{3}, -3\Delta
			_{2},-2\Delta _{2}-\Delta _{3},-\Delta _{2}-2\Delta _{3},   \nonumber \\
			& -3\Delta _{3}), \\
			\mathbf{C}_{[3]}^{(3)} =&{\footnotesize {\ \setlength{\arraycolsep}{1.2pt} \left( 
					\begin{array}{c|cc|ccc|cccc}
						\sqrt{3}g_{1} & g_{2} & g_{3} & 0 & 0 & 0 & 0 & 0 & 0 & 0 \\ \hline
						0 & \sqrt{2}g_{1} & 0 & \sqrt{2}g_{2} & g_{3} & 0 & 0 & 0 & 0 & 0 \\ 
						0 & 0 & \sqrt{2}g_{1} & 0 & g_{2} & \sqrt{2}g_{3} & 0 & 0 & 0 & 0 \\ \hline
						0 & 0 & 0 & g_{1} & 0 & 0 & \sqrt{3}g_{2} & g_{3} & 0 & 0 \\ 
						0 & 0 & 0 & 0 & g_{1} & 0 & 0 & \sqrt{2}g_{2} & \sqrt{2}g_{3} & 0 \\ 
						0 & 0 & 0 & 0 & 0 & g_{1} & 0 & 0 & g_{2} & \sqrt{3} g_{3}%
					\end{array}
					\right) }}  \nonumber \\
			=&{\small {\left( 
					\begin{array}{c|c|c|c}
						\mathbf{M1}_{[3]}^{(3)} & \mathbf{\tilde{M}1}_{[3]}^{(3)} & \mathbf{0} & \mathbf{0} \\ \hline
						\mathbf{0} & \mathbf{M2}_{[3]}^{(3)} &  \mathbf{\tilde{M}2}_{[3]}^{(3)} & \mathbf{0} \\ \hline
						\mathbf{0} & \mathbf{0} & \mathbf{M3}_{[3]}^{(3)} &  \mathbf{\tilde{M}3}_{[3]}^{(3)}%
					\end{array}
					\right) } .}  \label{C33}
		\end{align}%
	\end{subequations}
		
	The rank of $\mathbf{C}_{[3]}^{(3)}$ is six, and in the case of $\Delta _{1}=\Delta _{2}=\Delta _{3}=\Delta $, $\mathbf{L}_{[3]}^{(3)}$ becomes a ten-dimensional degenerate lower-state subspace, so there are four groups of linearly dependent vectors, which correspond to four dark states. The dark states can be derived as
	\begin{subequations}
		\begin{align}
			\left\vert D_{[3]}^{(3)}(1)\right\rangle   =&-g_{2}^{3}\left\vert g,3,0,0\right\rangle +%
			\sqrt{3}g_{2}^{2}g_{1}\left\vert g,2,1,0\right\rangle\notag \\
			& -\sqrt{3}	g_{2}g_{1}^{2}\left\vert g,1,2,0\right\rangle +g_{1}^{3}\left\vert
			g,0,3,0\right\rangle,  \\
			\left\vert D_{[3]}^{(3)}(2)\right\rangle   =&-\sqrt{3}g_{2}^{2}g_{3}\left\vert
			g,3,0,0\right\rangle +2g_{2}g_{3}g_{1}\left\vert g,2,1,0\right\rangle
			\notag \\
			&+g_{2}^{2}g_{1}\left\vert g,2,0,1\right\rangle-g_{3}g_{1}^{2}\left\vert
			g,1,2,0\right\rangle 	\notag \\
			& -\sqrt{2}g_{2}g_{1}^{2}\left\vert g,1,1,1\right\rangle+g_{1}^{3}\left\vert g,0,2,1\right\rangle,  \\
			\left\vert D_{[3]}^{(3)}(3)\right\rangle   =&-\sqrt{3}g_{2}g_{3}^{2}\left\vert
			g,3,0,0\right\rangle +g_{3}^{2}g_{1}\left\vert g,2,1,0\right\rangle
			\notag \\
			&+2g_{2}g_{3}g_{1}\left\vert g,2,0,1\right\rangle-\sqrt{2}%
			g_{3}g_{1}^{2}\left\vert g,1,1,1\right\rangle \notag \\
			& -g_{2}g_{1}^{2}\left\vert g,1,0,2\right\rangle+g_{1}^{3}\left\vert
			g,0,1,2\right\rangle , \\
			\left\vert D_{[3]}^{(3)}(4)\right\rangle=& -g_{3}^{3}\left\vert g,3,0,0\right\rangle -%
			\sqrt{3}g_{3}^{2}g_{1}\left\vert g,2,0,1\right\rangle  \notag \\
			&-\sqrt{3}g_{3}g_{1}^{2}\left\vert g,1,0,2\right\rangle+g_{1}^{3}\left\vert
			g,0,0,3\right\rangle .
		\end{align}\label{D33}%
	\end{subequations}
	
	Note that the above dark states are neither orthogonal nor normalized. Based on Eqs.~(\ref{D33}), however, a set of orthogonal dark states can be obtained using the Gram-Schmidt orthogonalization. Here we do not show the orthogonalized dark states to maintain conciseness. 
	
	\subsection{The dark states in the $\boldsymbol{n}$-excitation subspace}
	
	In the $n$-excitation subspace [Fig.~\ref{threemode}(d)], the numbers of the upper and lower states are, respectively, $n(n+1)/2$ and $(n+1)(n+2)/2$, which can also be represented by the combinatorial numbers $C_{n+1}^{2}$ and $	C_{n+2}^{2}$. Similarly, we write all the upper and lower states in order: $\{\left\vert	e,n-1,0,0\right\rangle ,$ $...,$ $ \left\vert e,n-1-s_{2}^{\prime},s_{2}^{\prime}-s_{3}^{\prime},s_{3}^{\prime}\right\rangle,$ $..., $ $\left\vert e,0,0,n-1\right\rangle ,$ $\left\vert g,n,0,0\right\rangle ,$ $...,$ $\left\vert g,n-s_{2},s_{2}-s_{3},s_{3}\right\rangle ,$ $...,$ $ \left\vert g,0,0,n\right\rangle \}$ with $s_{2}^{\prime}\in [0,n-1]$, $s_{3}^{\prime}\in [0,s_{2}^{\prime}]$, $s_{2}\in [0,n]$, and $s_{3}\in [0,s_{2}]$. By defining the basis vectors corresponding to these basis states as $( 1,0,0,0,...0) ^{T}$, $(0,1,0,0,...0) ^{T}$, ..., and $( 0,0,0,...0,1) ^{T}$, we can express the Hamiltonian $ \tilde{H}_{[3]}$ in the $n$-excitation subspace as
	\begin{equation}
		\tilde{H}_{[3]}^{(n)}=\left( 
		\begin{array}{c|c}
			(\mathbf{U}_{[3]}^{(n)} )_{C_{n+1}^{2} \times C_{n+1}^{2}}& (\mathbf{C}_{[3]}^{(n)})_{C_{n+1}^{2} \times C_{n+2}^{2}} \\ \hline
			\left(\mathbf{C}_{[3]}^{ (n)}\right)^{\dag }_{C_{n+2}^{2} \times C_{n+1}^{2}} &( \mathbf{L}_{[3]}^{(n)})_{C_{n+2}^{2}\times C_{n+2}^{2}}
		\end{array}
		\right) ,
	\end{equation}%
	where we introduce these submatrices	
	\begin{subequations}
		\begin{align}
			\mathbf{U}_{[3]}^{(n)} =&\text{diag}(-( n-1) \Delta _{1},...,-( n-1-s_{2}^{\prime})\Delta _{1}-(s_{2}^{\prime}-s_{3}^{\prime})\Delta _{2}\notag\\
			&-s_{3}^{\prime}\Delta _{3},...,-( n-1) \Delta _{3}), \\
			\mathbf{L}_{[3]}^{(n)} =&\text{diag}(-n\Delta _{1},...,-( n-s_{2}) \Delta _{1}-(s_{2}-s_{3})\Delta
			_{2}-s_{3}\Delta _{3},\notag\\
			&...,-n\Delta _{3}),\\
			\mathbf{C}_{[3]}^{(n)}=&{\left( 
			\begin{array}{c|c|c|c|c|c}
				\mathbf{M1}_{[3]}^{(n)} &  \mathbf{\tilde{M}1}_{[3]}^{(n)} & \mathbf{0} &... & \mathbf{0} & \mathbf{0} \\ \hline
				\mathbf{0} & \mathbf{M2}_{[3]}^{(n)} &\mathbf{\tilde{M}2}_{[3]}^{(n)}&...&  \mathbf{0} & \mathbf{0} \\  \hline
				\mathbf{0} &	\mathbf{0} & \mathbf{M3}_{[3]}^{(n)}  & ...&\mathbf{0} & \mathbf{0} \\ \hline
				... & ...& ... & ...  & ...& ... \\ \hline
				\mathbf{0} & \mathbf{0} & \mathbf{0} &...& \mathbf{\tilde{M}(n-1)}_{[3]}^{(n)} &\mathbf{0}\\\hline
				\mathbf{0} & \mathbf{0} & \mathbf{0} &... & \mathbf{Mn}_{[3]}^{(n)} & \mathbf{\tilde{M}n}_{[3]}^{(n)}
			\end{array}
			\right)}.  \label{c3n}
		\end{align}%
	\end{subequations}
	In Eq.~(\ref{c3n}), the submatrices are defined by
		\begin{subequations}
		\begin{align}
			&\mathbf{M(s_{2}^{\prime}+1)}_{[3]}^{(n)} = g_{1}\sqrt{n-s_{2}^{\prime}} \mathds{1}_{(s_{2}^{\prime}+1)}, \\
		&\mathbf{\tilde{M}(s_{2}^{\prime}+1)}_{[3]}^{(n)} \notag \\
		=&\small{{\left( 
					\begin{array}{ccccccc}
						\sqrt{s_{2}^{\prime}+1}g_{2} & g_{3} & 0 & 0 & 0 & 0 \\ 
						0 & ... & ... & 0 & 0 & 0  \\ 
						0 & 0 & \sqrt{s_{2}^{\prime}+1-s_{3}^{\prime}}g_{2} & \sqrt{s_{3}^{\prime}+1}g_{3}  & 0 & 0 \\ 
						0 & 0 & 0 & ... & ... & 0 \\ 
						0 & 0 & 0  & 0 & g_{2} & \sqrt{s_{2}^{\prime}+1}g_{3}%
					\end{array}
					\right)}},
		\end{align}%
	\end{subequations}
	where $\mathds{1}_{(s_{2}^{\prime}+1)}$ denotes the $(s_{2}^{\prime}+1)\times (s_{2}^{\prime}+1)$ identity matrix. Similarly, we can prove that the rank of the matrix $\mathbf{C}_{[3]}^{(n)}$ is $C_{n+1}^{2}$. 
	
	In the case of $\Delta_{1}=\Delta _{2}=\Delta _{3}=\Delta $, the number of dark states is equal to the difference between the column and row numbers of the coupling matrix, hence there are $n+1$ dark states. To obtain these linearly dependent vectors, we need to find the null space of the matrix $\mathbf{C}_{[3]}^{(n)}$. To this end, we solve the equation $\mathbf{C}_{[3]}^{(n)}\mathbf{x}=\mathbf{0}$ with $\mathbf{x}=(\mathbf{x}_{1},\mathbf{x}_{2},...,\mathbf{x}_{n},\mathbf{x}_{n+1})^{T}$. Based on the above equation and Eq.~(\ref{c3n}), we can obtain the relation
	\begin{equation}
		\mathbf{Mj}_{[3]}^{(n)}\mathbf{x}_{j}+\mathbf{\tilde{M}j}_{[3]}^{(n) }\mathbf{x}_{j+1}=\mathbf{0},\hspace{0.3cm}j=1,2,...,n,
		\label{Recurrence relation}
	\end{equation}%
	where $\mathbf{x}_{j}$ is the $j$-dimensional vector. We take $\mathbf{x}_{ n+1 }=( 0,0,1_{p},...,0,0) ^{T}$ (the corresponding state vector is $ \left\vert g,0,n-s_{3},s_{3}\right\rangle$ with $s_{2}=n$) and label the position as $p=s_{3}+1$, we can obtain
	\begin{eqnarray}
		\mathbf{x}_{j}&=&\sum_{k_{2},k_{3}}\frac{( -1) ^{k_{1}}\sqrt{
				k_{1}!A_{n-s_{3}}^{k_{2}}A_{s_{3}}^{k_{3}}}	}{ k_{2}!k_{3}!}g_{2}^{k_{2}}g_{3}^{k_{3}} \notag \\
			&&\times g_{1}^{n-k_{1}}	\left\vert g,k_{1},n-s_{3}-k_{2},s_{3}-k_{3}\right\rangle ,\label{v_n-p_3}
	\end{eqnarray}%
	where $k_{1}=n+1-j\in [1,n]$, and $k_{2}$ and $k_{3}$ satisfy the relation $k_{2}+k_{3}=k_{1}$ with $k_{2}\in[0,n-s_{3}]$ and $k_{3}\in[0,s_{3}]$. In Eq.~(\ref{v_n-p_3}), $A_{m}^{m^{\prime}}$ is the number of permutation defined by $A_{m}^{m^{\prime}}=m!/(m-m^{\prime})!$. Hence, we get the null space of the
	matrix, which corresponds to the dark states of the system, 
	\begin{eqnarray}
		\left\vert D_{[3]}^{(n)}(p)\right\rangle &=&\sum_{k_{1}=0}^{n}\sum_{k_{2},k_{3}}
		\frac{( -1) ^{k_{1}}\sqrt{
				k_{1}!A_{n-s_{3}}^{k_{2}}A_{s_{3}}^{k_{3}}}	}{ k_{2}!k_{3}!}g_{2}^{k_{2}}g_{3}^{k_{3}}\notag \\
			&&\times g_{1}^{n-k_{1}}\left\vert
		g,k_{1},n-s_{3}-k_{2},s_{3}-k_{3}\right\rangle ,\label{Da}
	\end{eqnarray}%
 and since $s_{3}\in [0,n]$, there are $n+1$ dark states. 	
	If we further introduce the mixing angle $\theta_{3} $ by $\tan \theta_{3} =g_{3}/g_{1}$, the dark states can be represented by mixing angles $\theta_{2} $ and $\theta_{3} $,
	\begin{eqnarray}
		\left\vert D_{[3]}^{(n)}(p)\right\rangle &=&\sum_{k_{1}=0}^{n}\sum_{k_{2},k_{3}}C_{k_{2},k_{3}}^{\prime}\sin ^{k_{2}} (\theta _{2})\cos^{n-s_{3}-k_{2}} (\theta _{2})\sin ^{k_{3}}
		(\theta_{3})
		\nonumber \\
		&&\times \cos ^{s_{3}-k_{3}}(\theta _{3})   \left\vert
		g,k_{1},n-s_{3}-k_{2},s_{3}-k_{3}\right\rangle,
	\end{eqnarray}%
	with $C_{k_{2},k_{3}}^{\prime}=( -1) ^{k_{1}}{\sqrt{
			k_{1}!A_{n-s_{3}}^{k_{2}}A_{s_{3}}^{k_{3}}}}/({k_{2} !k_{3}!})$.

	\subsection{The relationship between the dark states and dark modes}
	
	For the three-mode JC model, there also exists the dark-mode effect when the three field modes are degenerate. Concretely, for the case $\omega _{1}=\omega _{2}=\omega _{3}=\omega _{c}$, we can introduce one bright mode $(a_{3+})$ and two dark modes $(a_{2-},a_{3-})$ for the three-mode JC model as
	\begin{subequations}
		\begin{align}
			a_{3+}& =\frac{1}{\mathcal{N}_{[3]}^{(1)}}(g_{1}a_{1}+g_{2}a_{2}+g_{3}a_{3}), \\
			a_{2-}& =\frac{1}{\mathcal{N}_{[2]}^{(1)}}(g_{2}a_{1}-g_{1}a_{2}), \\
			a_{3-}& =\frac{1}{\mathcal{N}_{[2]}^{(1)}\mathcal{N}_{[3]}^{(1)}}
			\left[g_{3}(g_{1}a_{1}+g_{2}a_{2})-\mathcal{N}_{[2]}^{(2)} a_{3}\right].  
		\end{align}\label{a+1-2-}%
	\end{subequations}
	It can be shown that these modes satisfy the relation $[a_{r},a_{r^{\prime}}^{\dag}]=\delta_{rr^{\prime}}$ for $r,r^{\prime}=3+,2-,\text{ and }3-$. In the representation of these modes, the Hamiltonian of the three-mode JC model can be expressed as
	\begin{eqnarray}
		H_{[3]}&=&\frac{\omega _{0}}{2}\sigma _{z}+\omega _{c}( a_{3+}^{\dag
		}a_{3+}+a_{2-}^{\dag }a_{2-}+a_{3-}^{\dag }a_{3-}) \notag \\
&&		+\mathcal{N}_{[3]}^{(1)}( \sigma _{-}a_{3+}^{\dag }+a_{3+}\sigma_{+}) ,  \label{H_3_n}
	\end{eqnarray}%
	with the relationship  $a_{3+}^{\dag
		}a_{3+}+a_{2-}^{\dag }a_{2-}+a_{3-}^{\dag }a_{3-}=a_{1}^{\dag}a_{1}+a_{2}^{\dag}a_{2}+a_{3}^{\dag}a_{3}$ in the case of $\omega _{1}=\omega _{2}=\omega _{3}=\omega _{c}$. It can be confirmed that the dark modes $a_{2-}$ and $a_{3-}$ are decoupled from both the mode $a_{3+}$ and the atom. Similarly, the states with all excitations in the dark modes are dark states. Therefore, for the case of total excitation number operator $\mathrm{\hat{N}}_{[3]}=n$, there are $n+1$ combinations: $\left\vert 0\right\rangle _{a_{3+}}\left\vert m_{2}\right\rangle _{a_{2-}}\left\vert m_{3}\right\rangle _{a_{3-}}$ with $m_{2}+m_{3}=n$ and $0\le m_{2},m_{3} \le n $. If we arrange these $n+1$ dark states in columns, we obtain a $C_{n+1}^{2}\times ( n+1) $ matrix, and we denote it as $\mathbf{B}$. Since the states $\left\vert 0\right\rangle _{a_{3+}}\left\vert m_{2}\right\rangle _{a_{2-}}\left\vert m_{3}\right\rangle _{a_{3-}}$ are orthogonal to each other, the corresponding matrix $\mathbf{B}$ is of full rank. 
	
		\begin{figure*}[t]
		\centering\includegraphics[width=0.9\textwidth]{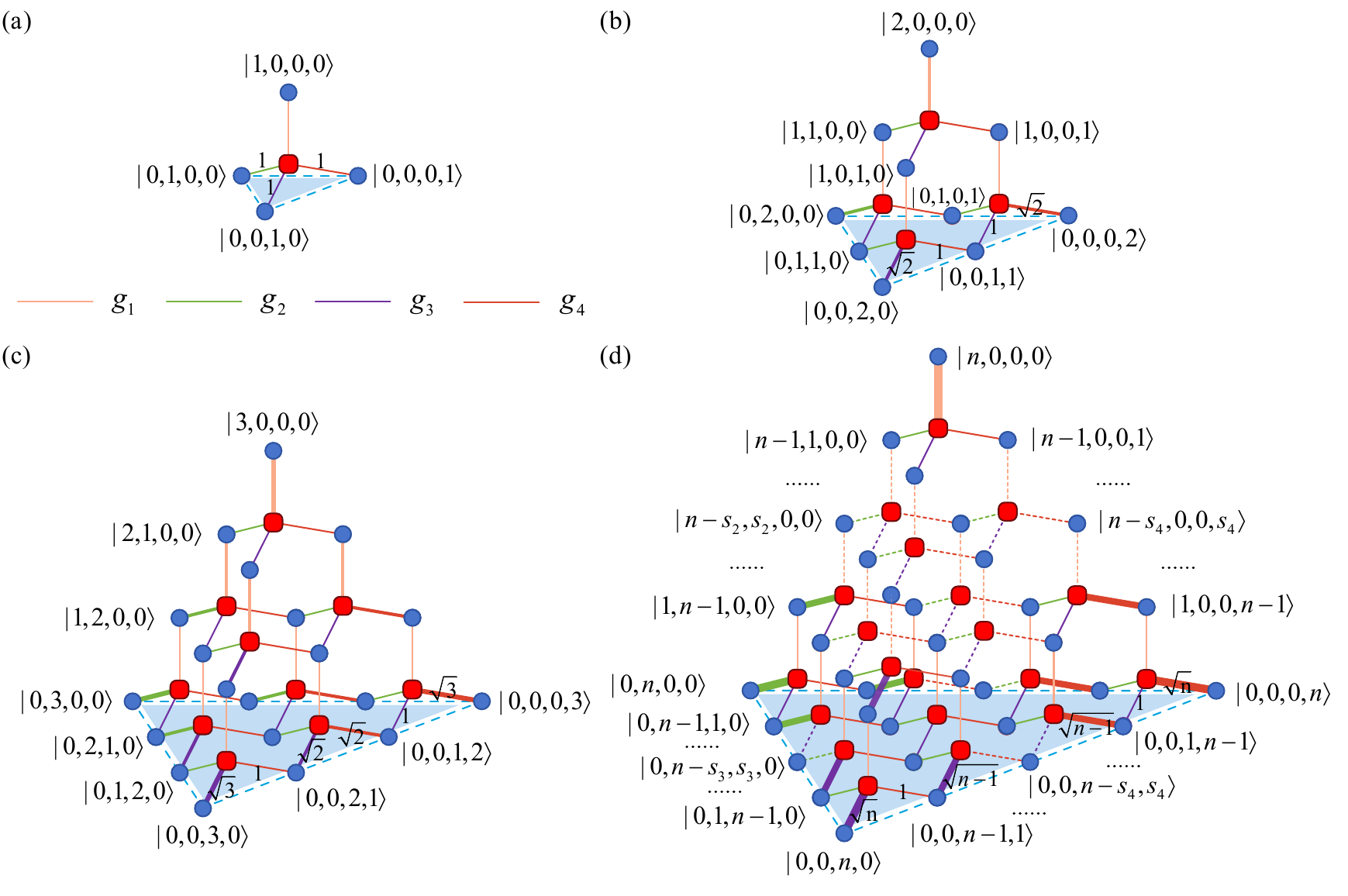}  
		\caption{The FSLs of the four-mode JC model restricted in the (a) single-, (b) double-, (c) triple-, and (d) $n$-excitation subspaces under the resonance condition $\Delta _{j=1-4}=\Delta $. The red squares (blue circles) denote the upper (lower) states of the system in the given-excitation-number subspaces. For clarity, we only label some of the necessary lower states in the FSLs.}
		\label{fourmode}
	\end{figure*}
	
	Similarly, the transpose of the null space of the matrix $\mathbf{C}_{[3]}^{(n)}$ can be written as an ${{n(n+1)}/{2}\times(n+1) }$ matrix, where the dark states are arranged in columns, and we record it as $\mathbf{A}$. Since we always take one free column with coefficient 1 and the other columns with coefficient 0 when solving $\mathbf{C}_{[3]}^{(n)}\mathbf{x}=\mathbf{0}$  [corresponding to $\mathbf{x}_{ n+1 }=( 0,0,1_{p},...,0,0) ^{T}$], then $\mathbf{x}_{ n+1 }$ are orthogonalized with each other for different $p$, hence these $n+1$ dark states are linearly independent, and the corresponding matrix $\mathbf{A}$ is of full rank.
	The dark states obtained by two different methods are characterized by the matrices $\mathbf{A}$ (Fock states) and $\mathbf{B}$ (dark modes), respectively. By comparing the two matrices, we find that they have the same dimension and rank, therefore we can prove that these two matrices are equivalent. In other words, the column vectors of these two matrices $\mathbf{A}$ and $\mathbf{B}$ span the same vector space (dark-state subspace).
	In fact, the matrix $\mathbf{B}$ is the result of the Gram-Schmidt orthogonalization of the vectors in matrix $\mathbf{A}$, which corresponds to the QR decomposition of $\mathbf{A}$. Consequently, the relation between $\mathbf{A}$ and $\mathbf{B}$ can be expressed as $\mathbf{A=BR}$, and the matrix $\mathbf{R}$ in this system takes the form as 
	\begin{equation}
		\mathbf{R}=\left(
		\begin{array}{cccc}
			r_{11} &r_{12} & ... & r_{1(n+1)}   \\ 
			0 & r_{22} & ... & r_{2(n+1)} \\ 
			\vdots  & \vdots  & \ddots  & \vdots    \\ 
			0 & 0 & ... & r_{(n+1)(n+1)}\\
		\end{array}	\right), 
	\end{equation}%
where $r_{kk}$ is the length of the columns after orthogonalization and $r_{ik}$ is the projection length of the $\mathbf{A}(:,k)$ in the direction of orthogonal vector $\mathbf{B}(:,i)$ with $i,k=1,2,...,n+1$, and $i \neq k$~\cite{Leon2014Applications}.

	Below, we calculate the dark states in the cases of single-excitation subspace and a general $n$-excitation subspace.	
	In the single-excitation subspace, the single excitation could be in either the dark modes $a_{2-}$ or $a_{3-}$, then the dark states can be written as 
	\begin{subequations}
		\begin{align}
			\left\vert \tilde{D}_{[3]}^{(1)}(1)\right\rangle   =&\left\vert g\right\rangle \left\vert
			0\right\rangle _{a_{3+}}\left\vert 1\right\rangle _{a_{2-}}\left\vert
			0\right\rangle _{a_{3-}} =\left\vert g\right\rangle \left\vert
			0\right\rangle _{a_{3+}}a_{2-}^{\dag}\left\vert 0\right\rangle _{a_{2-}}\left\vert
			0\right\rangle _{a_{3-}}, \\
			\left\vert \tilde{D}_{[3]}^{(1)}(2)\right\rangle   =&\left\vert g\right\rangle \left\vert
			0\right\rangle _{a_{3+}}\left\vert 0\right\rangle _{a_{2-}}\left\vert
			1\right\rangle _{a_{3-}} =\left\vert g\right\rangle \left\vert
			0\right\rangle _{a_{3+}}\left\vert 0\right\rangle _{a_{2-}}a_{3-}^{\dag}\left\vert
			0\right\rangle _{a_{3-}}.  
		\end{align}%
	\end{subequations}
	Based on Eqs.~(\ref{a+1-2-}), we can confirm that the above dark states are the same with the results given in Eqs.~(\ref{D1D2}), which confirms the relation between the dark modes and dark states.
	
	In the $n$-excitation subspace, the $n$ excitations are stored in the two modes $a_{2-}$ and $a_{3-}$, and there are $n+1$ different cases: $\left\vert m_{2}\right\rangle _{a_{2-}}\left\vert m_{3}\right\rangle _{a_{3-}}$ for $m_{2}+m_{3}=n$ and $0\le m_{2},m_{3} \le n $. Therefore, the dark states can be written as
	\begin{eqnarray}
			\left\vert \tilde{D}_{[3]}^{(n)}(m_{3}+1)\right\rangle & =&\left\vert g\right\rangle \left\vert
			0\right\rangle _{a_{3+}}\left\vert m_{2}\right\rangle _{a_{2-}}\left\vert
			m_{3}\right\rangle _{a_{3-}} \notag \\
			&=&	\frac{a_{2-}^{\dag m_{2}}a_{3-}^{\dag m_{3}}}{\sqrt{m_{2}!m_{3}!}}\left\vert g\right\rangle \left\vert
			0\right\rangle _{a_{3+}}\left\vert 0\right\rangle _{a_{2-}}\left\vert
			0\right\rangle _{a_{3-}} 	.\label{D1D2D3-n}%
	\end{eqnarray}%
	 Based on the above discussions, we see that both the two methods can be used to find the dark states in the three-mode JC model.
	
	\section{Dark states in the four-mode JC model}\label{sec5}
	
	In this section, we extend the above method to find the dark states in the FSLs for the four-mode JC model, which is described by the Hamiltonian $\tilde{H}_{[4]}$ [Eq.~(\ref{Hamiltonian2}) for $N=4$].

	\subsection{The dark states in the single-excitation subspace}
	In the single-excitation subspace [Fig.~\ref{fourmode}(a)], the basis states
	are $\{\left\vert e,0,0,0,0\right\rangle $, $\left\vert
	g,1,0,0,0\right\rangle $, $\left\vert g,0,1,0,0\right\rangle $, $\left\vert
	g,0,0,1,0\right\rangle $, $\left\vert g,0,0,0,1\right\rangle \}$, and there
	is one upper state $\left\vert e,0,0,0,0\right\rangle $ and four lower
	states $\{\left\vert g,1,0,0,0\right\rangle $, $\left\vert
	g,0,1,0,0\right\rangle ,$ $\left\vert g,0,0,1,0\right\rangle $, $\left\vert
	g,0,0,0,1\right\rangle \}$. We define the basis vectors: $\left\vert
	e,0,0,0,0\right\rangle =( 1,0,0,0,0) ^{T}$, $\left\vert
	g,1,0,0,0\right\rangle =( 0,1,0,0,0) ^{T}$, $\left\vert
	g,0,1,0,0\right\rangle =( 0,0,1,0,0) ^{T}$, $\left\vert
	g,0,0,1,0\right\rangle =( 0,0,0,1,0) ^{T}$, and $\left\vert
	g,0,0,0,1\right\rangle =( 0,0,0,0,1) ^{T}$. Then the Hamiltonian $%
	\tilde{H}_{[4]}$ in the single-excitation subspace can be expressed as 
	\begin{equation}
		\tilde{H}_{[4]}^{(1)}=\left( 
		\begin{array}{c|cccc}
			0 & g_{1} & g_{2} & g_{3} & g_{4} \\ \hline
			g_{1} & -\Delta _{1} & 0 & 0 & 0 \\ 
			g_{2} & 0 & -\Delta _{2} & 0 & 0 \\ 
			g_{3} & 0 & 0 & -\Delta _{3} & 0 \\ 
			g_{4} & 0 & 0 & 0 & -\Delta _{4}%
		\end{array}%
		\right) .
	\end{equation}%
	Here, the coupling matrix is a row vector with three groups of linear dependence when we choose the first column as the pivot column. In the case of $\Delta _{j=1-4}=\Delta$, there are three dark states  
	\begin{subequations}
		\begin{align}
			\left\vert D_{[4]}^{(1)}(1)\right\rangle &=\frac{g_{2}\left\vert g,1,0,0,0\right\rangle -g_{1}\left\vert
				g,0,1,0,0\right\rangle}{\sqrt{g_{2}^{2}+g_{1}^{2}}}  , \\
			\left\vert D_{[4]}^{(1)}(2)\right\rangle &=\frac{
				g_{3}\left\vert g,1,0,0,0\right\rangle -g_{1}\left\vert
				g,0,0,1,0\right\rangle }{\sqrt{g_{3}^{2}+g_{1}^{2}}} , \\
			\left\vert D_{[4]}^{(1)}(3)\right\rangle &=\frac{g_{4}\left\vert g,1,0,0,0\right\rangle -g_{1}\left\vert
				g,0,0,0,1\right\rangle}{\sqrt{g_{4}^{2}+g_{1}^{2}}}			  .
		\end{align}\label{D_4_1}%
	\end{subequations}
	With the Gram-Schmidt orthogonalization, we can obtain three orthogonalized dark states as 
	\begin{subequations}
		\begin{align}
			\left\vert \tilde{D}_{[4]}^{(1)}(1)\right\rangle   =&\frac{1}{\mathcal{N}_{[2]}^{(1)}}(g_{2}\left\vert
			g,1,0,0,0\right\rangle -g_{1}\left\vert g,0,1,0,0\right\rangle ), \\
			\left\vert \tilde{D}_{[4]}^{(1)}(2)\right\rangle   =&\frac{1}{\mathcal{N}_{[2]}^{(1)}\mathcal{N}_{[3]}^{(1)}}%
			(g_{3}g_{1}\left\vert g,1,0,0,0\right\rangle +g_{2}g_{3}\left\vert
			g,0,1,0,0\right\rangle \notag  \\
			&-\mathcal{N}_{[2]}^{(2)}\left\vert g,0,0,1,0\right\rangle ), \\
			\left\vert \tilde{D}_{[4]}^{(1)}(3)\right\rangle   =&\frac{1}{\mathcal{N}_{[3]}^{(1)}\mathcal{N}_{[4]}^{(1)}}%
			(g_{4}g_{1}\left\vert g,1,0,0,0\right\rangle +g_{2}g_{4}\left\vert
			g,0,1,0,0\right\rangle  \notag \\
			&+g_{3}g_{4}\left\vert g,0,0,1,0\right\rangle -\mathcal{N}_{[3]}^{(2)}\left\vert
			g,0,0,0,1\right\rangle ),
		\end{align} \label{D1D2D3_4}%
	\end{subequations}
	where we introduce the normalization constant $\mathcal{N}_{[4]}^{(1)}=(g_{1}^{2}+g_{2}^{2}+g_{3}^{2}+g_{4}^{2})^{1/2}	$. 
	
	\subsection{The dark states in the double-excitation subspace}
	In the double-excitation subspace [Fig.~\ref{fourmode}(b)], the basis states are $\{\left\vert e,1,0,0,0\right\rangle $, $\left\vert e,0,1,0,0\right\rangle $, $\left\vert e,0,0,1,0\right\rangle $, $\left\vert 	e,0,0,0,1\right\rangle $, $\left\vert g,2,0,0,0\right\rangle$, $\left\vert 	g,1,1,0,0\right\rangle$, $\left\vert g,1,0,1,0\right\rangle $, $\left\vert 	g,1,0,0,1\right\rangle$, $\left\vert g,0,2,0,0\right\rangle $, $\left\vert 	g,0,1,1,0\right\rangle $, $\left\vert g,0,1,0,1\right\rangle $, $\left\vert g,0,0,2,0\right\rangle $, $\left\vert g,0,0,1,1\right\rangle $, $\left\vert 	g,0,0,0,2\right\rangle \}$, and there are four upper states associated with atomic excited state $\left\vert e\right\rangle$ and ten lower states associated with atomic ground state $\left\vert g\right\rangle$. We define the basis vectors corresponding to these basis states as $( 1,0,0,0,...0) ^{T}$, $( 0,1,0,0,...0) ^{T}$, ..., and $( 0,0,0,...0,1) ^{T}$. The Hamiltonian $\tilde{H}_{[4]}$ in the 	double-excitation subspace can be expressed as 
	\begin{equation}
		\tilde{H}_{[4]}^{(2)}=\left( 
		\begin{array}{c|c}
			\mathbf{U}_{[4]}^{(2)} & \mathbf{C}_{[4]}^{(2)} \\ \hline
			\left(\mathbf{C}_{[4]}^{ (2)}\right)^{\dag } & \mathbf{L}_{[4]}^{(2)}%
		\end{array}
		\right) ,
	\end{equation}%
	where these submatrices are introduced as  
	\begin{subequations}
		\begin{align}
			\mathbf{U}_{[4]}^{(2)} =&\text{diag}(-\Delta _{1},-\Delta _{2},-\Delta _{3}, -\Delta
			_{4}), \\
			\mathbf{L}_{[4]}^{(2)} =&\text{diag}(-2\Delta _{1},-\Delta _{1}-\Delta _{2},-\Delta
			_{1}-\Delta _{3},-\Delta _{1}-\Delta _{4}, -2\Delta _{2},  \nonumber \\
			&-\Delta _{2}-\Delta _{3},-\Delta _{2}-\Delta _{4}, -2\Delta _{3},-\Delta
			_{3}-\Delta _{4},-2\Delta _{4}), \\
			\mathbf{C}_{[4]}^{(2)}  =&{\left( 
			\begin{array}{c|ccc|cccccc}
				\sqrt{2}g_{1} & g_{2} & g_{3} & g_{4} & 0 & 0 & 0 & 0 & 0 & 0 \\ \hline
				0 & g_{1} & 0 & 0 & \sqrt{2}g_{2} & g_{3} & g_{4} & 0 & 0 & 0 \\ 
				0 & 0 & g_{1} & 0 & 0 & g_{2} & 0 & \sqrt{2}g_{3} & g_{4} & 0 \\ 
				0 & 0 & 0 & g_{1} & 0 & 0 & g_{2} & 0 & g_{3} & \sqrt{2}g_{4}%
			\end{array}
			\right)}  \nonumber \\
			=&{\left( 
			\begin{array}{c|c|c}
				\mathbf{M1}_{[4]}^{(2)} & \mathbf{\tilde{M}1}_{[4]}^{(2)} & \mathbf{0} \\ \hline
				\mathbf{0} & \mathbf{M2}_{[4]}^{(2)} & \mathbf{\tilde{M}2}_{[4]}^{(2)}
			\end{array}
			\right)} .
		\end{align}%
	\end{subequations}
	Since $\mathbf{C}_{[4]}^{(2)}$ is a $4\times 10$ matrix, following the previous analyses we know that there are six dark states. Based on the null space of the coupling matrix $\mathbf{C}_{[4]}^{(2)}$, we can obtain these dark states
	\begin{subequations}
		\begin{align}
			\left\vert  D_{[4]}^{(2)}(1)\right\rangle   =&g_{2}^{2}\left\vert g,2,0,0,0\right\rangle
			-\sqrt{2}g_{2}g_{1}\left\vert g,1,1,0,0\right\rangle  \notag \\
			&+g_{1}^{2}\left\vert
			g,0,2,0,0\right\rangle , \\
			\left\vert  D_{[4]}^{(2)}(2)\right\rangle   =&\sqrt{2}g_{2}g_{3}\left\vert
			g,2,0,0,0\right\rangle -g_{3}g_{1}\left\vert g,1,1,0,0\right\rangle
			\notag \\
			&-g_{2}g_{1}\left\vert g,1,0,1,0\right\rangle+g_{1}^{2}\left\vert
			g,0,1,1,0\right\rangle , \\
			\left\vert  D_{[4]}^{(2)}(3)\right\rangle  =&\sqrt{2}g_{2}g_{4}\left\vert
			g,2,0,0,0\right\rangle -g_{4}g_{1}\left\vert g,1,1,0,0\right\rangle
			\notag \\
			&	-g_{2}g_{1}\left\vert g,1,0,0,1\right\rangle+g_{1}^{2}\left\vert
			g,0,1,0,1\right\rangle,  \\
			\left\vert  D_{[4]}^{(2)}(4)\right\rangle   =&g_{3}^{2}\left\vert g,2,0,0,0\right\rangle
			-\sqrt{2}g_{3}g_{1}\left\vert g,1,0,1,0\right\rangle  \notag \\
			&+g_{1}^{2}\left\vert
			g,0,0,2,0\right\rangle , \\
			\left\vert  D_{[4]}^{(2)}(5)\right\rangle   =&\sqrt{2}g_{3}g_{4}\left\vert
			g,2,0,0,0\right\rangle -g_{4}g_{1}\left\vert g,1,0,1,0\right\rangle
			\notag \\
			&-g_{3}g_{1}\left\vert g,1,0,0,1\right\rangle+g_{1}^{2}\left\vert
			g,0,0,1,1\right\rangle , \\
			\left\vert  D_{[4]}^{(2)}(6)\right\rangle  =&g_{4}^{2}\left\vert g,2,0,0,0\right\rangle
			-\sqrt{2}g_{4}g_{1}\left\vert g,1,0,0,1\right\rangle  \notag \\
			&+g_{1}^{2}\left\vert
			g,0,0,0,2\right\rangle .
		\end{align}%
	\end{subequations}
	Note that these dark states are neither orthogonal nor normalized, and a set of orthogonal dark states can be obtained by using the Gram-Schmidt orthogonalization. Here we do not show the orthogonalized dark states for keeping concise.

	\subsection{The dark states in the triple-excitation subspace}
	
	In the triple-excitation subspace [Fig.~\ref{fourmode}(c)], there are ten upper states associated with the atomic excited state $\left\vert e\right\rangle$ and twenty lower states associated with the atomic ground state $\left\vert g\right\rangle$. Similarly, by defining the basis vectors corresponding to these basis states as $( 1,0,0,0,...0) ^{T}$, $( 0,1,0,0,...0) ^{T}$, ..., and $( 0,0,0,...0,1) ^{T}$, we can obtain the Hamiltonian $ \tilde{H}_{[4]}$ in the triple-excitation subspace as
	\begin{equation}
		\tilde{H}_{[4]}^{(3)}=\left( 
		\begin{array}{c|c}
		\mathbf{U}_{[4]}^{(3)}	 & \mathbf{C}_{[4]}^{(3)} \\ \hline
		\left(\mathbf{C}_{[4]}^{ (3)}\right)^{\dag } & \mathbf{L}_{[4]}^{(3)}%
		\end{array}
		\right) ,
	\end{equation}%
	where we introduce these submatrices
	\begin{subequations}
		\begin{align}
			\mathbf{U}_{[4]}^{(3)} =&\text{diag}(-2\Delta _{1},-\Delta _{1}-\Delta _{2},-\Delta
			_{1}-\Delta _{3},-\Delta _{1}-\Delta _{4}, -2\Delta _{2},  \nonumber \\
			&  -\Delta _{2}-\Delta _{3}, -\Delta _{2}-\Delta _{4}, -2\Delta _{3} ,-\Delta _{3}-\Delta _{4},-2\Delta _{4}), \\
			\mathbf{L}_{[4]}^{(3)}=& \text{diag}(-3\Delta _{1},-2\Delta _{1}-\Delta
			_{2},-2\Delta _{1}-\Delta _{3},-2\Delta _{1}-\Delta _{4},-\Delta _{1}  \notag
			\\
			& -2\Delta _{2},-\Delta _{1}-\Delta _{2}-\Delta _{3},-\Delta _{1}-\Delta
			_{2}-\Delta _{4},-\Delta _{1}-2\Delta _{3},  \notag \\
			& -\Delta _{1}-\Delta _{3}-\Delta _{4},-\Delta _{1}-2\Delta _{4},-3\Delta
			_{2},-2\Delta _{2}-\Delta _{3},-2\Delta _{2}  \notag \\
			& -\Delta _{4},-\Delta _{2}-2\Delta _{3},-\Delta _{2}-\Delta _{3}-\Delta
			_{4},-\Delta _{2}-2\Delta _{4},-3\Delta _{3},  \notag \\
			& -2\Delta _{3}-\Delta _{4},-\Delta _{3}-2\Delta _{4},-3\Delta _{4}), \\
			\mathbf{C}_{[4]}^{(3)} =& \left( 
			\begin{array}{c|c|c|c}
				\mathbf{M1}_{[4]}^{(3)} & \mathbf{\tilde{M}1}_{[4]}^{ (3)} & \mathbf{0} & \mathbf{0} \\ \hline
				\mathbf{0} & \mathbf{M2}_{[4]}^{(3)} & \mathbf{\tilde{M}2}_{[4]}^{ (3)} & \mathbf{0} \\ \hline
				\mathbf{0} & \mathbf{0} & \mathbf{M3}_{[4]}^{(3)} & \mathbf{\tilde{M}3}_{[4]}^{ (3)}
			\end{array}%
			\right)	\label{c43}.
		\end{align}	%
	\end{subequations}
    In Eq.~(\ref{c43}), the submatrices are defined by
	\begin{subequations}
		\begin{align}
			\mathbf{M1}_{[4]}^{ (3)} =&\sqrt{3}g_{1} \mathds{1}_{1}, \hspace{0.1cm}\mathbf{M2}_{[4]}^{ (3)}= \sqrt{2}g_{1} \mathds{1}_{3 }, \hspace{0.1cm}\mathbf{M3}_{[4]}^{ (3)}= g_{1} \mathds{1}_{6 }, \\
			\mathbf{\tilde{M}1}_{[4]}^{ (3)} =&( 
			\begin{array}{ccc}
				g_{2} & g_{3} & g_{4}
			\end{array}	) ,	\\
			\mathbf{\tilde{M}2}_{[4]}^{ (3)} =&\left( 
			\begin{array}{c|c|c}
				\sqrt{2}g_{2} &\mathbf{\tilde{N}1}_{[4]}^{ (3)} & \mathbf{0}  \\ \hline
				\mathbf{0} & g_{2}\mathds{1}_{2} & \mathbf{\tilde{N}2}_{[4]}^{ (3)}
			\end{array}
			\right) , \\
			\mathbf{\tilde{M}3}_{[4]}^{ (3)}  =&\left( 
			\begin{array}{c|c|c|c}
				\sqrt{3}g_{2} &\mathbf{\tilde{N}1}_{[4]}^{ (3)} & \mathbf{0} & \mathbf{0}  \\ \hline
				\mathbf{0} & \sqrt{2}g_{2}\mathds{1}_{2} & \mathbf{\tilde{N}2}_{[4]}^{ (3)} &  \mathbf{0} \\ \hline
				\mathbf{0} & \mathbf{0} & g_{2}\mathds{1}_{3 } & \mathbf{\tilde{N}3}_{[4]}^{ (3)}
			\end{array}
			\right) , \\
			\mathbf{\tilde{N}1}_{[4]}^{ (3)} =&( 
			\begin{array}{cc}
				g_{3} & g_{4}
			\end{array}	) ,	 \\
			\mathbf{\tilde{N}2}_{[4]}^{ (3)} =&\left( 
			\begin{array}{ccc}
				\sqrt{2}g_{3} & g_{4} &  0 \\
				0	& g_{3} & \sqrt{2}g_{4} 
			\end{array}	\right) ,\\
			\mathbf{\tilde{N}3}_{[4]}^{ (3)} =&\left( 
			\begin{array}{cccccccccc}
				\sqrt{3}g_{3} & g_{4} & 0 & 0 \\ 
				0 & 	\sqrt{2}g_{3} & \sqrt{2}g_{4} & 0 \\ 
				0 & 0& g_{3} & \sqrt{3}g_{4}%
			\end{array}
			\right).
		\end{align}	%
	\end{subequations}
	Since $\mathbf{C}_{[4]}^{(3)}$ is a $10\times 20$ matrix, we know that there are ten dark states in the case $\Delta _{j=1-4}=\Delta$. We can also find the specific form of the dark states by solving the null space of the coupling matrix,
	\begin{subequations}
		\begin{align}
			\left\vert  D_{[4]}^{(3)}(1)\right\rangle  =&-g_{2}^{3}\left\vert g,3,0,0,0\right\rangle
			+\sqrt{3}g_{2}^{2}g_{1}\left\vert g,2,1,0,0\right\rangle   \notag \\
			& -\sqrt{3}g_{2}g_{1}^{2}\left\vert g,1,2,0,0\right\rangle
			+g_{1}^{3}\left\vert g,0,3,0,0\right\rangle , \\
			\left\vert D_{[4]}^{(3)}(2)\right\rangle  =&-\sqrt{3}g_{2}^{2}g_{3}\left\vert
			g,3,0,0,0\right\rangle +2g_{2}g_{3}g_{1}\left\vert g,2,1,0,0\right\rangle  
			\notag \\
			& +g_{2}^{2}g_{1}\left\vert g,2,0,1,0\right\rangle -g_{3}g_{1}^{2}\left\vert
			g,1,2,0,0\right\rangle  \notag \\
			& -\sqrt{2}g_{2}g_{1}^{2}\left\vert g,1,1,1,0\right\rangle
			+g_{1}^{3}\left\vert g,0,2,1,0\right\rangle , \\
			\left\vert D_{[4]}^{(3)}(3)\right\rangle  =&-\sqrt{3}g_{2}^{2}g_{4}\left\vert
			g,3,0,0,0\right\rangle +2g_{2}g_{4}g_{1}\left\vert g,2,1,0,0\right\rangle  
			\notag \\
			& +g_{2}^{2}g_{1}\left\vert g,2,0,0,1\right\rangle -g_{4}g_{1}^{2}\left\vert
			g,1,2,0,0\right\rangle \notag \\
			& -\sqrt{2}g_{2}g_{1}^{2}\left\vert g,1,1,0,1\right\rangle
			+g_{1}^{3}\left\vert g,0,2,0,1\right\rangle , \\
			\left\vert D_{[4]}^{(3)}(4)\right\rangle  =&-\sqrt{3}g_{2}g_{3}^{2}\left\vert
			g,3,0,0,0\right\rangle +g_{3}^{2}g_{1}\left\vert g,2,1,0,0\right\rangle  
			\notag \\
			& +2g_{2}g_{3}g_{1}\left\vert g,2,0,1,0\right\rangle -\sqrt{2}%
			g_{3}g_{1}^{2}\left\vert g,1,1,1,0\right\rangle \notag \\
			& -g_{2}g_{1}^{2}\left\vert g,1,0,2,0\right\rangle +g_{1}^{3}\left\vert
			g,0,1,2,0\right\rangle , \\
			\left\vert D_{[4]}^{(3)}(5)\right\rangle =& {-\sqrt{6}g_{2}g_{3}g_{4}\left%
			\vert g,3,0,0,0\right\rangle +\sqrt{2}g_{3}g_{4}g_{1}\left\vert
			g,2,1,0,0\right\rangle }  \notag \\
			& { +\sqrt{2} g_{2} g_{4} g_{1}\left\vert
			g,2,0,1,0\right\rangle  +\sqrt{2} g_{2} g_{3} %
				g_{1}\left\vert g,2,0,0,1\right\rangle}   \notag \\
			& -g_{4}g_{1}^{2}\left\vert g,1,1,1,0\right\rangle -g_{3}g_{1}^{2}\left\vert
			g,1,1,0,1\right\rangle   \notag \\
			& -g_{2}g_{1}^{2}\left\vert g,1,0,1,1\right\rangle +g_{1}^{3}\left\vert
			g,0,1,1,1\right\rangle , \\
			\left\vert D_{[4]}^{(3)}(6)\right\rangle  =&-\sqrt{3}g_{2}g_{4}^{2}\left\vert
			g,3,0,0,0\right\rangle +g_{4}^{2}g_{1}\left\vert g,2,1,0,0\right\rangle  
			\notag \\
			& +2g_{2}g_{4}g_{1}\left\vert g,2,0,0,1\right\rangle -\sqrt{2}%
			g_{4}g_{1}^{2}\left\vert g,1,1,0,1\right\rangle  \notag \\
			& -g_{2}g_{1}^{2}\left\vert g,1,0,0,2\right\rangle +g_{1}^{3}\left\vert
			g,0,1,0,2\right\rangle , \\
			\left\vert D_{[4]}^{(3)}(7)\right\rangle  =&-g_{3}^{3}\left\vert g,3,0,0,0\right\rangle
			+\sqrt{3}g_{3}^{2}g_{1}\left\vert g,2,0,1,0\right\rangle  \notag \\
			& -\sqrt{3}g_{3}g_{1}^{2}\left\vert g,1,0,2,0\right\rangle
			+g_{1}^{3}\left\vert g,0,0,3,0\right\rangle , \\
			\left\vert D_{[4]}^{(3)}(8)\right\rangle  =&-\sqrt{3}g_{3}^{2}g_{4}\left\vert
			g,3,0,0,0\right\rangle +2g_{3}g_{4}g_{1}\left\vert g,2,0,1,0\right\rangle \notag \\
			& +g_{3}^{2}g_{1}\left\vert g,2,0,0,1\right\rangle -g_{4}g_{1}^{2}\left\vert
			g,1,0,2,0\right\rangle \notag \\
			& -\sqrt{2}g_{3}g_{1}^{2}\left\vert g,1,0,1,1\right\rangle
			+g_{1}^{3}\left\vert g,0,0,2,1\right\rangle , \\
			\left\vert D_{[4]}^{(3)}(9)\right\rangle  =&-\sqrt{3}g_{3}g_{4}^{2}\left\vert
			g,3,0,0,0\right\rangle +g_{4}^{2}g_{1}\left\vert g,2,0,1,0\right\rangle \notag \\
			& +2g_{3}g_{4}g_{1}\left\vert g,2,0,0,1\right\rangle -\sqrt{2}%
			g_{4}g_{1}^{2}\left\vert g,1,0,1,1\right\rangle \notag \\
			& -g_{3}g_{1}^{2}\left\vert g,1,0,0,2\right\rangle +g_{1}^{3}\left\vert
			g,0,0,1,2\right\rangle , \\
			\left\vert D_{[4]}^{(3)}(10)\right\rangle  =&-g_{4}^{3}\left\vert
			g,3,0,0,0\right\rangle +\sqrt{3}g_{4}^{2}g_{1}\left\vert
			g,2,0,0,1\right\rangle  \notag \\
			& -\sqrt{3}g_{4}g_{1}^{2}\left\vert g,1,0,0,2\right\rangle
			+g_{1}^{3}\left\vert g,0,0,0,3\right\rangle .
		\end{align}\label{d43}%
	\end{subequations}
A set of orthogonal dark states can be obtained based on Eqs.~(\ref{d43}) using the Gram-Schmidt orthogonalization. Here, we do not show the orthogonalized dark states for keeping concise.
	
	\subsection{The dark states in the $\boldsymbol{n}$-excitation subspace}	
	In the $n$-excitation subspace [Fig.~\ref{fourmode}(d)], there are $C_{n+2}^{3}$ upper states associated with the atomic excited state $\left\vert e\right\rangle$ and $C_{n+3}^{3}$ lower states associated with the atomic ground state $\left\vert g\right\rangle$. Similarly, we write all the upper and lower states in order: $\{\left\vert e,n-1,0,0,0\right\rangle ,$ $...,$ $ \left\vert e,n-1-s_{2}^{\prime},s_{2}^{\prime}-s_{3}^{\prime},s_{3}^{\prime}-s_{4}^{\prime},s_{4}^{\prime} \right\rangle,$ $..., $ $\left\vert e,0,0,0,n-1\right\rangle ,$ $\left\vert g,n,0,0,0\right\rangle ,$ $...,$ $\left\vert
	g,n-s_{2},s_{2}-s_{3},s_{3}-s_{4},s_{4}\right\rangle ,$ $...,$ $ \left\vert g,0,0,0,n\right\rangle\}$ with $s_{2}^{\prime}\in [0,n-1]$, $s_{3}^{\prime}\in [0,s_{2}^{\prime}]$, $s_{4}^{\prime}\in [0,s_{3}^{\prime}]$, $s_{2}\in [0,n]$, $s_{3}\in [0,s_{2}]$, and $s_{4}\in [0,s_{3}]$. We define the basis vectors corresponding to these basis states as $( 1,0,0,0,...0) ^{T}$, $(0,1,0,0,...0) ^{T}$, ..., and $( 0,0,0,...0,1) ^{T}$, then we can obtain the Hamiltonian $ \tilde{H}_{[4]}$ in the $n$-excitation subspace as
	\begin{equation}
		\tilde{H}_{[4]}^{(n)}=\left( 
		\begin{array}{c|c}
			(\mathbf{U}_{[4]}^{(n)})_{C_{n+2}^{3} \times C_{n+2}^{3}} & (\mathbf{C}_{[4]}^{(n)})_{C_{n+2}^{3} \times C_{n+3}^{3}} \\ \hline
			\left(\mathbf{C}_{[4]}^{ (n)}\right)^{\dag }_{C_{n+3}^{3} \times C_{n+2}^{3}} & (\mathbf{L}_{[4]}^{(n)})_{C_{n+3}^{3} \times C_{n+3}^{3}}
		\end{array}
		\right) ,
	\end{equation}%
	where we introduce these submatrices	
	\begin{subequations}
		\begin{align}
			\mathbf{U}_{[4]}^{(n)} =&\text{diag}(-( n-1) \Delta _{1},...,-( n-1-s_{2}^{\prime})
			\Delta _{1}-(s_{2}^{\prime}-s_{3}^{\prime})\Delta _{2}\notag\\
			&-(s_{3}^{\prime}-s_{4}^{\prime})\Delta _{3}-s_{4}^{\prime}\Delta _{4},...,-( n-1) \Delta _{4}), \\
			\mathbf{L}_{[4]}^{(n)} =&\text{diag}(-n\Delta _{1},...,-( n-s_{2}) \Delta _{1}-(s_{2}-s_{3})\Delta_{2}-(s_{3}-s_{4})\Delta_{3}\notag\\
			&-s_{4}\Delta _{4},...,-n\Delta _{4}), \\
			\mathbf{C}_{[4]}^{(n)}=&{\left( 
			\begin{array}{c|c|c|c|c|c}
				\mathbf{M1}_{[4]}^{(n)} &  \mathbf{\tilde{M}1}_{[4]}^{(n)} & \mathbf{0} &... & \mathbf{0} & \mathbf{0} \\ \hline
				\mathbf{0} & \mathbf{M2}_{[4]}^{(n)} &\mathbf{\tilde{M}2}_{[4]}^{(n)}&... &  \mathbf{0} & \mathbf{0} \\  \hline
				\mathbf{0} &	\mathbf{0} & \mathbf{M3}_{[4]}^{(n)}  & ...&\mathbf{0} & \mathbf{0} \\ \hline
				... & ...& ... & ...  & ...& ... \\ \hline
				\mathbf{0} & \mathbf{0} & \mathbf{0} &...& \mathbf{\tilde{M}(n-1)}_{[4]}^{(n)} &\mathbf{0}\\\hline
				\mathbf{0} & \mathbf{0} & \mathbf{0} &... & \mathbf{Mn}_{[4]}^{(n)} & \mathbf{\tilde{M}n}_{[4]}^{(n)}
			\end{array}
			\right)}. \label{c4n}
		\end{align}	%
	\end{subequations}
	In Eq.~(\ref{c4n}), the submatrices are defined by  
\begin{subequations}
	\begin{align}
		&\small{\mathbf{M(s_{2}^{\prime}+1)}_{[4]}^{(n)}}=
		\small{ g_{1}\sqrt{n-s_{2}^{\prime}} \mathds{1}_{{(s_{2}^{\prime}+1)(s_{2}^{\prime}+2)}/{2} }}, \\
		&\small{\mathbf{\tilde{M}(s_{2}^{\prime}+1)}_{[4]}^{ (n)}}\notag \\
		=&\footnotesize{	\left( 
			\begin{array}{cccccc}
				\small{\mathbf{N1}_{[4]}^{(n)}} & \small{\mathbf{\tilde{N}1}_{[4]}^{(n)}} & \mathbf{0} & \mathbf{0} & \mathbf{0} & \mathbf{0} \\ 
				\mathbf{0} & ... & ... & \mathbf{0} & \mathbf{0} & \mathbf{0} \\ 
				\mathbf{0} & \mathbf{0} &\small{\mathbf{N(s_{3}^{\prime}+1)}_{[4]}^{(n)}}  & \small{\mathbf{\tilde{N}(s_{3}^{\prime}+1)}_{[4]}^{(n)}} & \mathbf{0} & \mathbf{0} \\ 
				\mathbf{0} & \mathbf{0} & \mathbf{0} & ... & ... & \mathbf{0} \\ 
				\mathbf{0} & \mathbf{0} & \mathbf{0} & \mathbf{0} & \small{\mathbf{N(s_{3}^{\prime}+1)}_{[4]}^{(n)}} &  \small{\mathbf{\tilde{N}(s_{3}^{\prime}+1)}_{[4]}^{(n)}}
			\end{array}	\right)} , \\
		&\small{\mathbf{N(s_{3}^{\prime}+1)}_{[4]}^{(n)}}=
		\small{ g_{2}\sqrt{n-s_{3}^{\prime}} \mathds{1}_{{(s_{3}^{\prime}+1) }}}, \\
		&\small{\mathbf{\tilde{N}(s_{3}^{\prime}+1)}_{[4]}^{(n)}}\notag \\
		=&\small{\left( 
			\begin{array}{cccccc}
				\sqrt{s_{3}^{\prime}+1}g_{3} & g_{4} & 0 & 0 & 0 & 0  \\ 
				0 & ... & ... & 0 & 0 & 0   \\ 
				0 & 0 & \small{\sqrt{s_{3}^{\prime}+1-s_{4}^{\prime}}g_{3}} & \sqrt{s_{4}^{\prime}+1}g_{4} & 0 & 0  \\ 
				0 & 0 & 0 & ... & ... & 0 \\ 
				0 & 0 & 0 & 0 & g_{3} & \sqrt{s_{3}^{\prime}+1}g_{4}  
			\end{array}
			\right).}
	\end{align}%
\end{subequations}
	In the case of $\Delta _{j=1-4}=\Delta$, we can solve the null space of the matrix $\mathbf{C}_{[4]}^{(n)}$ with the similar method in Eq.~(\ref{Recurrence relation}).
	Note that here $\mathbf{x}_{j} $ is a ${j(j+1)}/{2}$-dimensional vector.
	
	We take $\mathbf{x}_{n+1} =(0,0,...,1_{p},...,0,0) ^{T}$ (the corresponding state vector is $\left\vert g,0,n-s_{3},s_{3}-s_{4},s_{4}\right\rangle$ with $s_{2}=n$) and label the position as $p=s_{3}(s_{3}+1)/2+s_{4}+1$.
    Then we obtain 
	\begin{eqnarray}
		\mathbf{x}_{j}&=&\sum_{k_{2},k_{3},k_{4}}\frac{( -1)
			^{k_{1}}\sqrt{k_{1}!A_{n-s_{3}}^{k_{2}}A_{s_{3}-s_{4}}^{k_{3}}A_{s_{4}}^{k_{4}}}}{k_{2}!k_{3}!k_{4}!}g_{2}^{k_{2}}g_{3}^{k_{3}}g_{4}^{k_{4}}\notag \\
			&&\times g_{1}^{n-k_{1}}\left\vert
		g,k_{1},n-s_{3}-k_{2},s_{3}-s_{4}-k_{3},s_{4}-k_{4}\right\rangle ,\label{v_n-p_4}
	\end{eqnarray}%
	where $k_{1}=n+1-j\in [1,n]$. In addition, $k_{2}$, $k_{3}$, and $k_{4}$ satisfy the relation $k_{2}+k_{3}+k_{4}=k_{1}$ with $k_{2}\in[0,n-s_{3}]$, $k_{3}\in[0,s_{3}-s_{4}]$, and $k_{4}\in[0,s_{4}]$.
	Hence, we get the null space of the matrix, which determines the dark states of the system, 
		\begin{equation}
		\small{	\left\vert D_{[4]}^{(n)}(p)\right\rangle
			=\sum_{k_{1}=0}^{n}\sum_{k_{2},k_{3},k_{4}}C_{k_{2},k_{3},k_{4}}\left\vert
			g,k_{1},n-s_{3}-k_{2},s_{3}-s_{4}-k_{3},s_{4}-k_{4}\right\rangle }, \label{Dabc}
	\end{equation}%
	with 
	\begin{equation}
		C_{k_{2},k_{3},k_{4}}=\frac{( -1)
			^{k_{1}}\sqrt{k_{1}!A_{n-s_{3}}^{k_{2}}A_{s_{3}-s_{4}}^{k_{3}}A_{s_{4}}^{k_{4}}}}{k_{2}!k_{3}!k_{4}!}g_{1}^{n-k_{1}}g_{2}^{k_{2}}g_{3}^{k_{3}}g_{4}^{k_{4}}.
	\end{equation} %
By introducing the mixing angle $\theta_{4}  $ by $\tan \theta_{4} =g_{4}/g_{1}$, the dark states can be represented by mixing angles $\theta_{2} $, $\theta_{3} $, and $\theta_{4} $,
	\begin{eqnarray}
	  {	\left\vert D_{[4]}^{(n)}(p) \right\rangle}
		&=&	  {\sum_{k_{1}=0}^{n}\sum_{k_{2},k_{3},k_{4}} C_{k_{2},k_{3},k_{4}}^{\prime}   
		  \sin  ^{k_{2}}(\theta_{2}) \cos^{n-s_{3}-k_{2}} (\theta_{2})  } \nonumber \\
		 	&&  {\times \sin ^{k_{3}}(\theta_{3} ) \cos	^{s_{3}-s_{4}-k_{3}} (\theta_{3})    \sin  ^{k_{4}}(\theta_{4})\cos  ^{s_{4}-k_{4}}(\theta_{4})  }\nonumber \\
		&&  {\times   
		\left\vert g,k_{1},n-s_{3}-k_{2},s_{3}-s_{4}-k_{3},s_{4}-k_{4}\right\rangle },
	\end{eqnarray}%
	with $ C_{k_{2},k_{3},k_{4}}^{\prime}={( -1)	^{k_{1}}\sqrt{k_{1}!A_{n-s_{3}}^{k_{2}}A_{s_{3}-s_{4}}^{k_{3}}A_{s_{4}}^{k_{4}}}}/({k_{2}!k_{3}!k_{4}!})$.

	\subsection{The relationship between the dark states and dark modes}
	
	For the four-mode JC model, there also exists the dark-mode effect when the four field modes are degenerate. Concretely, for the case $\omega _{j=1-4}=\omega _{c}$, we can introduce the bright mode and dark modes for the four-mode JC model as
	\begin{subequations}
		\begin{align}
			a_{4+}& =\frac{1}{\mathcal{N}_{[4]}^{(1)}}(g_{1}a_{1}+g_{2}a_{2}+g_{3}a_{3}+g_{4}a_{4}), \\
			a_{2-}& =\frac{1}{\mathcal{N}_{[2]}^{(1)}}(g_{2}a_{1}-g_{1}a_{2}), \\
			a_{3-}& =\frac{1}{\mathcal{N}_{[2]}^{(1)}\mathcal{N}_{[3]}^{(1)}}\left[g_{3}( g_{1}a_{1}+g_{2}a_{2}) -
			\mathcal{N}_{[2]}^{(2)} a_{3}\right], \\
			a_{4-}& =\frac{1}{\mathcal{N}_{[3]}^{(1)}\mathcal{N}_{[4]}^{(1)}}\left[g_{4}( g_{1}a_{1}+g_{2}a_{2}+g_{3}a_{3}) -\mathcal{N}_{[3]}^{(2)} a_{4}\right]	, 
		\end{align}\label{a+1-2-3-}%
	\end{subequations}
	which satisfy the relation $[a_{r},a_{r^{\prime}}^{\dag}]=\delta_{rr^{\prime}}$ for $r,r^{\prime}=4+,$ $2-,3-,\text{ and }4-$. In the representation of these modes, the Hamiltonian of the four-mode JC model can be expressed as
	\begin{equation}
		H_{[4]}=\frac{\omega _{0}}{2}\sigma _{z}+\omega _{c} \left(a_{4+}^{\dag
		}a_{4+}+\sum_{l=2}^{4}a_{l-}^{\dag }a_{l-} \right)+\mathcal{N}_{[4]}^{(1)}( \sigma _{-}a_{4+}^{\dag }+a_{4+}\sigma _{+}) , \label{H_4_n}
	\end{equation}%
	with the relationship  $a_{4+}^{\dag
		}a_{4+}+\sum_{l=2}^{4}a_{l-}^{\dag }a_{l-}=\sum_{j=1}^{4}a_{j-}^{\dag }a_{j-}$ in the case of $\omega _{j=1-4}=\omega _{c}$. We find the dark modes $a_{2-}$, $a_{3-}$, and $a_{4-}$ are decoupled from both the atom and the bright mode $a_{4+}$. In the $n$-excitation subspace, when $\mathrm{\hat{N}}_{[4]}=n $, there are $C_{n+2}^{2}$ basis states in the new representation: $\left\vert  0\right\rangle _{a_{4+}}\left\vert m_{2}\right\rangle _{a_{2-}}\left\vert m_{3}\right\rangle _{a_{3-}}\left\vert m_{4}\right\rangle _{a_{4-}}$ with $m_{2}+m_{3}+m_{4}=n$ and $0\le m_{2},m_{3} ,m_{4}\le n $. In this case, all the excitations are stored in the dark modes $( a_{2-},a_{3-},\text{ and }a_{4-}) $,
	which corresponds to $C_{n+2}^{2}$ dark states. Similarly, if we arrange these $C_{n+2}^{2}$ dark states in columns, we can get a matrix with dimension $ C_{n+2}^{3}\times C_{n+2}^{2}$, and we denote it as $\mathbf{B}$. Since $\left\vert 0\right\rangle_{a_{4+}}\left\vert m_{2}\right\rangle _{a_{2-}}\left\vert m_{3}\right\rangle _{a_{3-}}\left\vert m_{4}\right\rangle _{a_{4-}}$ are orthogonal to each other, the matrix consisting of these $C_{n+2}^{2}$ vectors is of full rank. We record the transpose of the null space of the matrix $\mathbf{C}_{[4]}^{(n)}$ as $\mathbf{A}$, namely, we arrange the $C_{n+2}^{2}$ dark states in Eq.~(\ref{Dabc}) in columns, and the matrix $\mathbf{A}$ is of full rank. Therefore, we can prove that these two matrices are equivalent, and the dark states obtained by the two different methods are consistent. 
	
In the following, we present the result in the single-excitation subspace and a general $n$-excitation subspace. In the single-excitation subspace, there are three dark states,
	\begin{subequations}
		\begin{align}
			\left\vert \tilde{D}_{[4]}^{(1)}(1)\right\rangle  =&\left\vert g\right\rangle \left\vert
			0\right\rangle _{a_{4+}}\left\vert 1\right\rangle _{a_{2-}}\left\vert
			0\right\rangle _{a_{3-}}\left\vert 0\right\rangle _{a_{4-}}\notag \\
			=&\left\vert g\right\rangle \left\vert
			0\right\rangle _{a_{4+}}a_{2-}^{\dag}\left\vert 0\right\rangle _{a_{2-}}\left\vert
			0\right\rangle _{a_{3-}}\left\vert 0\right\rangle _{a_{4-}}, \\
			\left\vert \tilde{D}_{[4]}^{(1)}(2)\right\rangle   =&\left\vert g\right\rangle \left\vert
			0\right\rangle _{a_{4+}}\left\vert 0\right\rangle _{a_{2-}}\left\vert
			1\right\rangle _{a_{3-}}\left\vert 0\right\rangle _{a_{4-}}\notag \\
			=& \left\vert g\right\rangle \left\vert
			0\right\rangle _{a_{4+}}\left\vert 0\right\rangle _{a_{2-}}a_{3-}^{\dag}\left\vert
			0\right\rangle _{a_{3-}}\left\vert 0\right\rangle _{a_{4-}}, \\
			\left\vert \tilde{D}_{[4]}^{(1)}(3)\right\rangle   =&\left\vert g\right\rangle \left\vert
			0\right\rangle _{a_{4+}}\left\vert 0\right\rangle _{a_{2-}}\left\vert
			0\right\rangle _{a_{3-}}\left\vert 1\right\rangle _{a_{4-}}\notag \\
			=&\left\vert g\right\rangle \left\vert
			0\right\rangle _{a_{4+}}\left\vert 0\right\rangle _{a_{2-}}\left\vert
			0\right\rangle _{a_{3-}}a_{4-}^{\dag}\left\vert 0\right\rangle _{a_{4-}} .
		\end{align} \label{D123-n}%
	\end{subequations}
	In terms of Eqs.~(\ref{a+1-2-3-}), we can show that the dark states given in Eqs.~(\ref{D123-n}) are identical to the results given in Eqs.~(\ref{D1D2D3_4}). This indicates that the normalized orthogonal dark states are consistent with the dark states obtained by the dark-mode method.
	
	In the $n$-excitation subspace, the $n$ excitations are stored in the three modes $a_{2-}$, $a_{3-}$, and $a_{4-}$, and there are $C_{n+2}^{2}$ different cases: $\left\vert m_{2}\right\rangle _{a_{2-}}\left\vert m_{3}\right\rangle _{a_{3-}}\left\vert m_{4}\right\rangle _{a_{4-}}$ for $m_{2}+m_{3}+m_{4}=n$ and $0\le m_{2},m_{3} ,m_{4}\le n $. Therefore, the dark states can be written as
	\begin{eqnarray}
			\left\vert \tilde{D}_{[4]}^{(n)}(p)\right\rangle  &=&\left\vert g\right\rangle \left\vert
			0\right\rangle _{a_{4+}}\left\vert m_{2}\right\rangle _{a_{2-}}\left\vert
			m_{3}\right\rangle _{a_{3-}} \left\vert m_{4}\right\rangle _{a_{4-}}\notag \\
			&=&	\frac{a_{2-}^{\dag m_{2}}a_{3-}^{\dag m_{3}}a_{4-}^{\dag m_{4}}}{\sqrt{m_{2}!m_{3}!m_{4}!}}\left\vert g\right\rangle \left\vert
			0\right\rangle _{a_{4+}}\left\vert 0\right\rangle _{a_{2-}}\left\vert
			0\right\rangle _{a_{3-}}  \left\vert 0\right\rangle _{a_{4-}}	, \label{D4n}
	\end{eqnarray}%
	with $p=(m_{3}+m_{4})(m_{3}+m_{4}+1)/2+m_{4}+1$. Based on Eqs.~(\ref{a+1-2-3-}) and (\ref{D4n}), we can confirm that the two methods can be used to find the dark states in the four-mode JC model.

	\section{Dark states in the \textit{N}-mode JC model}\label{sec6}
	In this section, we study the dark states for the $N$-mode JC model. We will present the dark states in the single-excitation subspace and a general $n$-excitation subspace.
	
	\subsection{The dark states in the single-excitation subspace}
	In the single-excitation subspace, the basis states are given by $\{\left\vert e,0,...,0,...,0\right\rangle ,$ $\left\vert g,1,0,...,0\right\rangle$, ..., $\left\vert g,0,...,0,1\right\rangle \}$, and there is one upper state associated with the atomic excited state $\left\vert e\right\rangle$ and $N$ lower states associated with the atomic ground state $\left\vert g\right\rangle$. We define the basis vectors: $\left\vert e,0,...,0,...,0\right\rangle =( 1,0,0,...,0) ^{T}$, $\left\vert g,1,0,...,0\right\rangle =( 0,1,0,...,0) ^{T}$, ..., and $\left\vert g,0,...,0,1\right\rangle =( 0,0,...,0,1) ^{T}$, then the Hamiltonian $\tilde{H}_{[N]}$ in the single-excitation subspace can be expressed as 
	\begin{equation}
		\tilde{H}_{[N]}^{(1)}=\left( 
		\begin{array}{c|cccccc}
			0 & g_{1} & g_{2} & \cdots  & g_{j} & \cdots  & g_{N} \\ \hline
			g_{1} & -\Delta _{1} & 0 & \cdots  & 0 & \cdots  & 0 \\ 
			g_{2} & 0 & -\Delta _{2} & \cdots  & 0 & \cdots  & 0 \\ 
			\vdots  & \vdots  & \vdots  & \ddots  & \vdots  & \cdots  & \vdots  \\ 
			g_{j} & 0 & 0 & \cdots  & -\Delta _{j} & \cdots  & 0 \\ 
			\vdots  & \vdots  & \vdots  & \cdots  & \vdots  & \ddots  & \vdots  \\ 
			g_{N} & 0 & 0 & \cdots  & 0 & \cdots  & -\Delta _{N}%
		\end{array}%
		\right) .
	\end{equation}%
	In the case of $\Delta _{j=1-N}=\Delta $, there are $N-1$ dark states  
	\begin{subequations}
		\begin{align}
			\left\vert D_{[N]}^{(1)}(1)\right\rangle & =\frac{g_{2}\left\vert
				g,1,0,...,0\right\rangle -g_{1}\left\vert g,0,1,0,...,0\right\rangle }{\sqrt{%
					g_{2}^{2}+g_{1}^{2}}}, \\
			\left\vert D_{[N]}^{(1)}(2)\right\rangle & =\frac{g_{3}\left\vert
				g,1,0,...,0\right\rangle -g_{1}\left\vert g,0,0,1,...,0\right\rangle }{\sqrt{%
					g_{3}^{2}+g_{1}^{2}}}, \\
			& ...  \notag \\
			\left\vert D_{[N]}^{(1)}(l^{\prime})\right\rangle & =\frac{g_{l^{\prime}+1}\left\vert
				g,1,0,...,0\right\rangle -g_{1}\left\vert g,...,1_{(l^{\prime}+1)},...,0\right\rangle }{\sqrt{%
					g_{l^{\prime}+1}^{2}+g_{1}^{2}}}, \\
			& ...  \notag \\
			\left\vert D_{[N]}^{(1)}(N-1)\right\rangle & =\frac{g_{N}\left\vert
				g,1,0,...,0\right\rangle -g_{1}\left\vert g,0,0,0,...,1\right\rangle }{\sqrt{%
					g_{N}^{2}+g_{1}^{2}}}.
		\end{align}%
	\end{subequations}
	where $l^{\prime}=1,2,...,N-1$ and the subscript of the element 1 denotes the position of the element 1. Note that the forms of the dark states are not unique. With the Gram-Schmidt orthogonalization, we can obtain the orthogonalized dark states as 
	\begin{subequations}
		\begin{align}
			\left\vert \tilde{D}_{[N]}^{(1)}(1)\right\rangle  =&\frac{1}{\mathcal{N}_{[2]}^{(1)}}( g_{2}\left\vert          g,1,0,...,0\right\rangle 		-g_{1}\left\vert g,0,1,...,0\right\rangle ) , \\
			\left\vert \tilde{D}_{[N]}^{(1)}(2)\right\rangle  =&\frac{1}{\mathcal{N}_{[2]}^{(1)}\mathcal{N}_{[3]}^{(1)}}%
			( g_{1}g_{3}\left\vert g,1,0,...,0\right\rangle  \notag\\ 
			&+g_{3}g_{2}\left\vert
			g,0,1,...,0\right\rangle \notag\\ 
			&
			-\mathcal{N}_{[2]}^{(2)} \left\vert
			g,0,0,1,...,0\right\rangle ) , \\
			... \notag\\  
			\left\vert \tilde{D}_{[N]}^{(1)}(l^{\prime})\right\rangle  =&\frac{1}{\mathcal{N}_{[l^{\prime}]}^{(1)}\mathcal{N}_{[l^{\prime}+1]}^{(1)}}(g_{l^{\prime}+1}g_{1}\left%
			\vert g,1,0,...,0\right\rangle  \notag \\
			&+g_{l^{\prime}+1}g_{2}\left\vert
			g,0,1,...,0\right\rangle +...\notag \\
			&+g_{l^{\prime}+1}g_{l^{\prime}}\left\vert g,0,...,1_{(l^{\prime})},...,0\right\rangle \notag \\ &-\mathcal{N}_{[l^{\prime}]}^{(2)}\left\vert g,0,..,1_{\left( l^{\prime}+1\right)
			},...,0\right\rangle  ), \\
			...\notag  \\ 
			\left\vert \tilde{D}_{[N]}^{(1)}(N-1)\right\rangle  =&\frac{1}{\mathcal{N}_{[N-1]}^{(1)}\mathcal{N}_{[N]}^{(1)}}(g_{N}g_{1}\left\vert
			g,1,0,...,0\right\rangle \notag \\
			&+g_{N}g_{2}\left\vert g,0,1,...,0\right\rangle+...  \notag  \\
			&+g_{N}g_{N-1}\left\vert g,0,...,1,0\right\rangle \notag\\
			&-\mathcal{N}_{[N]}^{(2)} \left\vert g,0,0,..,1\right\rangle ),
		\end{align} \label{DN}%
	\end{subequations} 
	where we introduce the normalization constant $\mathcal{N}_{[l^{\prime}]}^{(1)}=(g_{1}^{2}+g_{2}^{2}+g_{3}^{2}+...+g_{l^{\prime}}^{2})^{1/2}$.

	\subsection{Analyses of the dark states in the $\boldsymbol{n}$-excitation subspace}
	In this subsection, we present some analyses on the dark states in the $n$-excitation subspace. In principle, the number and form of these dark states can be derived using the same method. Below, we only present the number of these dark states because the forms of these dark states are too complicated.

	In the $n$-excitation subspace, the number of the upper and lower states can be represented by combination numbers: $C_{N+n-2}^{N-1} $ and $C_{N+n-1}^{N-1} $. We arrange the basis states in order, and by defining the basis vectors corresponding to these basis states as $( 1,0,0,0,...0) ^{T}$, $(0,1,0,0,...0) ^{T}$, ..., and $( 0,0,0,...0,1) ^{T}$, we can obtain the matrix corresponding to the Hamiltonian $\tilde{H}_{[N]}$ restricted in the $n$-excitation subspace. In this case, the submatrices in Eq.~(\ref{HNn}) can be written as  
		\begin{subequations}
		\begin{align}
			\mathbf{U}_{[N]}^{(n)} =&\text{diag}(-( n-1) \Delta _{1},-( n-2) \Delta _{1}-\Delta _{2},-( n-2) \Delta _{1}-\Delta _{3},...,\notag\\
			&-( n-1-s_{2}^{\prime}) \Delta _{1}-(s_{2}^{\prime}-s_{3}^{\prime})\Delta _{2}-...-(s_{n-1}^{\prime}-s_{n}^{\prime})\Delta _{N-1}\notag\\
			&-s_{n}^{\prime}\Delta _{N},...,-( n-1) \Delta _{N}), \\
			\mathbf{L}_{[N]}^{(n)} =&\text{diag}(-n \Delta _{1},-( n-1) \Delta _{1}-\Delta _{2},-( n-1) \Delta _{1}-\Delta _{3},...,\notag\\
			&-( n-s_{2}) \Delta _{1}-(s_{2}-s_{3})\Delta _{2}-...-(s_{n-1}-s_{n})\Delta _{N-1}\notag\\
			&-s_{n}\Delta _{N},...,- n \Delta _{N}), 
		\end{align}	%
	\end{subequations}
	and the matrix $\mathbf{C}_{[N]}^{(n)}$ is given in Eq.~(\ref{CNn}). The submatrices $ \mathbf{M(s_{2}^{\prime}+1)}_{[N]}^{(n)} $ and $ \mathbf{\tilde{M}(s_{2}^{\prime}+1)}_{[N]}^{(n)}$ can be obtained by calculating the matrix elements and the matrix $ \mathbf{M(s_{2}^{\prime}+1)}_{[N]}^{(n)}$ is a $C_{N+s_{2}^{\prime}-2}^{N-2}\times C_{N+s_{2}^{\prime}-2}^{N-2}$ identity matrix with coefficient $\sqrt{n-s_{2}^{\prime}}g_{1}$. However, the form of the matrix $ \mathbf{\tilde{M}(s_{2}^{\prime}+1)}_{[N]}^{(n)}$ is complicate and here we do not present the expression of $ \mathbf{\tilde{M}(s_{2}^{\prime}+1)^{}}_{[N]}^{(n)}$ for concise. In principle, the matrix $ \mathbf{\tilde{M}(s_{2}^{\prime}+1)}_{[N]}^{(n)}$ can be obtained by using computer codes.
	
    	\begin{table}[t!]
    	\caption{The number of the dark states in the $n$-excitation subspace for the $N$-mode JC model under various values of $N$ and $n$.}
    	\centering
    	\renewcommand{\arraystretch}{2.3} % 设置行高
    	\begin{tabular}{|c|c|c|c|c|}
    		\hline
    		&$n=1$ & 	$n=2$ & 	$n=3$ & $n$ \\	\hline
    		$N=2$&  $1$ & $1$ &$1$& $1$ \\ \hline
    		$N=3$ &  $2$ & $3$&  $4$&  $C_{n+1}^{1}$  \\ \hline
    		$N=4$ &$3$ & $6$&$10$ & $C_{n+2}^{2}$    \\ \hline
    		$N$   &$C_{N-1}^{N-2}$  & $C_{N}^{N-2}$&$C_{N+1}^{N-2}$&  $C_{N+n-2}^{N-2}$\\ \hline
    	\end{tabular}\label{table}
    \end{table}

    When $\Delta _{j=1-N}=\Delta $, the number of the dark states can be obtained as
	\begin{eqnarray}
		C_{N+n-1}^{N-1}-C_{N+n-2}^{N-1} =\frac{( N+n-2) !}{(N-2) ! n!}=C_{N+n-2}^{N-2}. \label{formula}
	\end{eqnarray}%
	For various values of $N$ and $n$, namely, in the $n$-excitation subspace of the $N$-mode JC model, the number of the dark states can be obtained by Eq.~(\ref{formula}). In Table~\ref{table}, we present the number of the dark states when $N=2,3,4$ and $n=1,2,3$.  We point out that the forms of these dark states can be obtained by solving the null space of the coupling matrix $\mathbf{C}_{[N]}^{(n)}$. For a realistic case, we can use numerical programs to calculate the coupling matrix. Then the null space can also be calculated numerically. The result in Eq.~(\ref{formula}) confirms those obtained in previous sections.
	
	\subsection{Expressing the dark states with the dark modes}
	
		For the $N$-mode JC model, there also exists the dark-mode effect when the $N$ field modes are degenerate. Concretely, for the case of $\omega _{j=1-N}=\omega _{c}$, we can introduce one bright mode and $N-1$ dark modes for the $N$-mode JC model as~\cite{huang2023dark}	
	\begin{subequations}
		\begin{align}
			a_{N+}=&\frac{1}{\mathcal{N}_{[N]}^{(1)}}(
			g_{1}a_{1}+g_{2}a_{2}+...+g_{N}a_{N}),  \\
			a_{l-}=&\frac{1}{\mathcal{N}_{[l-1]}^{(1)}\mathcal{N}_{[l]}^{(1)}}\left[g_{l}(
			g_{1}a_{1}+g_{2}a_{2}+...+g_{l-1}a_{l-1})		-\mathcal{N}_{[l-1]}^{(2)}a_{l}\right],
		\end{align} \label{al-}%
	\end{subequations}%
	which satisfy the relation $[a_{r},a_{r^{\prime}}^{\dag}]=\delta_{rr^{\prime}}$ for $r,r^{\prime}=N+,l-$ with $l=2,3,...,N$. In the representation of these modes, the Hamiltonian of the $N$-mode JC model can be expressed as
	\begin{equation}
	\small {	H_{[N]}=\frac{\omega _{0}}{2}\sigma _{z}+\omega _{c} \left(a_{N+}^{\dag
		}a_{N+}+\sum_{l=2}^{N}a_{l-}^{\dag }a_{l-} \right)+\mathcal{N}_{[N]}^{(1)}( \sigma _{-}a_{N+}^{\dag }+a_{N+}\sigma _{+}) }, \label{H_N_n}
	\end{equation}%
	with the relationship  $a_{N+}^{\dag
		}a_{N+}+\sum_{l=2}^{N}a_{l-}^{\dag }a_{l-}=\sum_{j=1}^{N}a_{j-}^{\dag }a_{j-}$ in the case of $\omega _{j=1-N}=\omega _{c}$. We find these dark modes $a_{l-}$ are decoupled from both the atom and the bright mode $a_{N+}$.
	When the total excitation number operator $\mathrm{\hat{N}}_{[N]}=n $, there are $C_{N+n-2}^{N-2}$ basis states in the new representation: $\left\vert 0\right\rangle _{a_{N+}}\left\vert m_{2}\right\rangle _{a_{2-}}\left\vert
	m_{3}\right\rangle _{a_{3-}}...\left\vert m_{N}\right\rangle _{a_{N-}}$ with $\sum_{l=2}^{N} m_{l}=n$ and $0\le m_{l=2-N} \le n $. In this case, the excitations are all stored in the dark modes $ a_{l-},$ which corresponds to $C_{N+n-2}^{N-2}$ dark states. 
	
	 Below, we calculate the dark states in the cases of single- and $n$-excitation subspaces.	
	In the single-excitation subspace, the single excitation could be in any one of the dark modes $a_{l-}$, then the dark states can be written as 
	\begin{eqnarray}
		\left\vert \tilde{D}_{[N]}^{(1)}(l^{\prime})\right\rangle 	&=&\left\vert g\right\rangle \left\vert
		0\right\rangle _{a_{N+}}\left\vert 0\right\rangle _{a_{2-}}\left\vert
		0\right\rangle _{a_{3-}}...\left\vert 1\right\rangle _{a_{l-}}...\left\vert 0\right\rangle _{a_{N-}}  \notag \\
		&=&\left\vert g\right\rangle \left\vert
		0\right\rangle _{a_{N+}}\left\vert 0\right\rangle _{a_{2-}}\left\vert
		0\right\rangle _{a_{3-}}...a_{l-}^{\dag}\left\vert 0\right\rangle _{a_{l-}}...\left\vert 0\right\rangle _{a_{N-}} , \label{Ddl}
	\end{eqnarray}%
	with $l^{\prime }=l-1$. Here, we used the relations in Eqs.~(\ref{al-}), and the results given in Eq.~(\ref{Ddl}) are consistent with the states in Eqs.~(\ref{DN}).
	
	In the $n$-excitation subspace, these $n$ excitations could be in these $N-1$ dark modes, then there are $C_{N+n-2}^{N-2}$ possible arrangements: $\left\vert m_{2}\right\rangle _{a_{2-}}\left\vert	m_{3}\right\rangle _{a_{3-}}...\left\vert m_{N}\right\rangle _{a_{N-}}$ for $\sum_{l=2}^{N} m_{l}=n$ and $0\le m_{l=2-N} \le n $, and the dark states can be expressed as 
		\begin{eqnarray}
			\left\vert \tilde{D}_{[N]}^{(n)}(p)\right\rangle 	&=&\left\vert g\right\rangle \left\vert	0\right\rangle _{a_{N+}}\left\vert m_{2}\right\rangle _{a_{2-}}\left\vert
			m_{3}\right\rangle _{a_{3-}}...\left\vert m_{N}\right\rangle _{a_{N-}}  \notag \\
			&=&\prod_{l=2}^{N} \frac{a_{l-}^{\dag m_{l}}}{\sqrt{m_{l}!}}\left\vert g\right\rangle \left\vert	0\right\rangle _{a_{N+}}	\left\vert0\right\rangle _{a_{2-}}
			\left\vert0\right\rangle _{a_{3-}}...\left\vert0\right\rangle _{a_{N-}},
	\end{eqnarray}%
	where  $p=1,2,...,C_{N+n-2}^{N-2}$. Based on the above discussions, we see that both the two methods can be used to find the dark states in the $N$-mode JC model.

	\section{Discussions and conclusions}\label{conclusion}
	
	Finally, we present some discussions on the possible experimental implementation of the multimode JC models, the possible experimental observation of the dark-state effect, the potential applications of the dark states, and the advantages of the FSLs and the arrowhead-matrix method. 
	
(i) Our discussions are based on the general multimode JC models. Therefore, our scheme should be implemented in the physical platforms that can be used to realize the multimode JC models. Namely, the candidate setups should be able to realize the couplings between a TLA and multiple bosonic modes, and there should be no couplings between these bosonic modes. The detunings between the TLA and these bosonic modes should be tunable, and we can choose the identical-detunings case to make sure the appearance of the dark states. Currently, the multimode JC models can be implemented with many physical platforms, such as cavity-QED systems~\cite{CQED:rmp2001,CQED2023,CQED2024}, circuit-QED systems~\cite{CiQED2010,CiQED2013,CiQED:prl2015,CiQED:prx2015,CiQED2020,CiQED:rmp2021,topology:science2022,CiQED:arx2024}, circuit-QAD systems~\cite{CiQED2024}, and trapped-ion systems~\cite{TRi:rmp2003,TRi2024}.
	
	(ii) In the multimode JC models, the dark states are the eigenstates of the Hamiltonian. However, we can find an interesting phenomenon: the forms of the dark states only depend on these coupling strengths $g_{j}$, while these eigenenergies corresponding to these dark states only depend on the detunings $\Delta_{j=1-N}=\Delta$, independent of these coupling strengths $g_{j}$. Therefore, this feature should be able to be observed from	the observables of the multimode JC models. Note that the dark-state effect has been detected from the occupations of the cavity modes in the cavity-QED system~\cite{CQED2024}.
		
	(iii) The dark-state effect has potential application in quantum optics and quantum information science. For example, the STIRAP technique has been used to achieve the perfect quantum state transfer and frequency conversion. For the cases with degenerate dark subspaces, both the STIRAP process and quantum state transfer have been discussed~\cite{STIRAPiDEG2001,STIRAPiDEG2008,STIRAPiDEG2013}. Therefore, the dark-state effect in the multimode JC models can enable quantum information processing in JC quantum networks, which will play an important role in quantum science and technology in the near future.

	(iv) Finally, we want to mention the advantages of the FSLs and the arrowhead-matrix method. Though our method is general and it works for any excitation subspaces, most considerations of the dark states focus on the low excitation case, because realistic quantum information and quantum optical experiments are typically associated with low excitations, namely in finite-dimensional spaces. In realistic physical systems, when weak driving and vacuum dissipation are considered, the system can still be treated as a low-excitation case under proper approximate conditions. In fact, when the driving is introduced, there will be no dark states. This is because either driving the atom or the cavity will promote the transition of the atomic ground state to the excited state, and thus the dark state can not exist.
	Note that, in the open-system case, we can still extend the concept of the FSLs and construct the FSLs in the Liouville space~\cite{LsiaX}. Moreover, the structure of the FSLs can be engineered by properly designing the coupling terms, which makes the construction of FSLs flexible~\cite{FSLtQO2023}.
	Therefore, the FSLs and the arrowhead-matrix method are powerful tools for analyzing the dark states. 
	Furthermore, the Fock-state representation is naturally useful to the discrete-variable systems, which are commonly considered in quantum computing and communication.

	In conclusion, we have studied the dark states in the FSLs for the multimode JC models by employing the arrowhead-matrix method. We have obtained the number and form of the dark states in fixed-excitation subspaces of the two-, three-, and four-mode JC models. We have also found that the dark-mode method can provide an alternative perspective to determine the form of the dark states in the $n$-excitation subspace. In particular, we have established the connection between the dark states and dark modes. 
	Finally, we have generalized the method to analyze the number of dark states in certain excitation-number subspaces for the $N$-mode JC model. It is found that the number of the dark states in the $n$-excitation subspace for the $N$-mode JC model is $C_{N+n-2}^{N-2}$. We have also discussed the implementation of quantum state transfer with the STIRAP mechanism based on the dark states in the FSLs.

	\begin{acknowledgments}
	The authors would like to thank Dr. Han Cai for helpful discussions on the connection between the dark states and the topological properties of the FSLs. J.-Q.L. was supported in part by National Natural Science Foundation of China (Grants No.~12175061, No.~12247105, No.~11935006, and No.~12421005), National Key Research and Development Program of China (Grant No.~2024YFE0102400), and Hunan Provincial Major Sci-Tech Program (Grant No.~2023ZJ1010). L.-M.K. is supported by the Natural Science Foundation of China (NSFC) (Grants No.~12247105, No.~12175060, No.~11935006, No.~12421005), Hunan Provincial Major Sci-Tech Program (Grant No.~2023ZJ1010), and  XJ-Lab key project (Grant No.~23XJ02001).
	\end{acknowledgments}

\end{document}